\documentclass[preprint,aps,showpacs,showkeys,tightenlines,superscriptaddress]{revtex4}
\usepackage[dvips]{graphicx}
\newcommand{\bfi}[1]{\mbox{\boldmath $#1$}}

\def \be{\begin{equation}}
\def \ee{\end{equation}}
\def \bea{\begin{eqnarray}}
\def \eea{\end{eqnarray}}
\begin{document}
\title{Structure and decay width of $\Theta^+$ in a 
one-gluon exchange model}
\author{H. Matsumura}
\email{hideki@nt.sc.niigata-u.ac.jp}
\affiliation{Graduate School of Science and Technology, Niigata University, Niigata 950-2181, Japan}
\author{Y. Suzuki}
\email{suzuki@nt.sc.niigata-u.ac.jp}
\affiliation{Department of Physics, Niigata University, Niigata
950-2181, Japan}

\begin{abstract}
The mass and decay width of the $\Theta^+(1540)$ with isospin 0 
are investigated in a constituent quark model comprising $uudd\bar{s}$ 
quarks. The resonance state for the $\Theta^+$ is identified as a 
stable solution in correlated basis calculations. 
With the use of a one-gluon exchange quark-quark 
interaction, the mass is calculated to be larger than 
2 GeV, increasing in order of the spin-parity, $\frac{1}{2}^-, \, 
\frac{3}{2}^-$  
and $\frac{1}{2}^+ (\frac{3}{2}^+)$, and only the $\frac{3}{2}^-$ state 
has a small width to the $nK^{*+}$ decay. 
If the calculated mass is shifted 
to around 100 MeV above the $N\!+\!K$ threshold, the $\Theta^+(1540)$ 
is possibly $\frac{1}{2}^+ (\frac{3}{2}^+)$ or $\frac{3}{2}^-$, 
though in the latter case it cannot decay to the $nK^{+}$ 
channel. In addition it is conjectured that other pentaquark state 
with different 
spin-parity exists below the $\Theta^+(1540)$. The structure 
of the $\Theta^+$ is discussed through the densities and two-particle 
correlation functions of 
the quarks and through the wave function decomposition to a 
baryon-meson model and a diquark-pair model. 
\end{abstract}

\maketitle
\section{Introduction}
\label{intro}

Quark model succeeds to classify the hundreds of baryons and mesons. 
A constituent quark model well reproduces the mass spectra 
of these hadrons using a colored interaction of one-gluon 
exchange (OGE) type together with a phenomenological confinement 
potential~\cite{rujula,al1}. There are of course those hadrons which do not 
fit the prediction of the constituent quark model, e.g., 
$N(1440)$ (Roper resonance) and $\Lambda(1405)$. Recently, 
motivated by a theoretical prediction for an exotic
baryon~\cite{diakonov}, some experimental groups have searched for it 
and reported the observation of $\Theta^+(1540)$ which has 
baryon number $+1$, strangeness $+1$ and a decay width of 
less than 25 MeV~\cite{nakano,stepanyan,barth}. Its minimal 
quark content is $uudd{\bar s}$, so it is often called a 
pentaquark baryon. A reason why the $\Theta^+$ attracts 
much interest is that it is truly exotic, its mass is  
about 100 MeV above the $N\!+\!K$ threshold, and its decay width 
appears to be very small nevertheless. 
The spin and parity of the $\Theta^+$ has not yet 
been known, nor even the existence of the $\Theta^+(1540)$ seems 
to be established experimentally~\cite{hicks,adamovich}. 

The recent controversial status of the $\Theta^+(1540)$ 
warrants a careful study of this system. From a theoretical 
side a calculation of the pentaquark mass and its decay 
width is useful and important.  
It was speculated that the $\Theta^+$ is a bound state of two 
highly correlated $ud$ pairs (diquark) and 
${\bar s}$~\cite{jaffe} or it is a coupled system of a color 
antisymmetric $ud$ diquark with a $ud{\bar s}$ 
triquark~\cite{karliner}. In these models the spin and 
parity of the $\Theta^+$ was assumed to be $J^P=\frac{1}{2}^+$ 
to account for its small decay width.

Some dynamical calculations of the mass of the $\Theta^+$ for 
different $J^P$ states have been performed in constituent quark 
models~\cite{takeuchi,enyo,paris,hiyama}. The quark model usually 
predicts a smaller mass for the negative parity $\Theta^+$ than 
for the positive parity $\Theta^+$. In fact the ground state of the 
$\Theta^+$ is $\frac{1}{2}^-$, but it 
was argued that the mass difference between the 
$\frac{1}{2}^{-}$ and $\frac{1}{2}^{+}$ states may become small 
or even be reversed 
for a certain type of flavor-spin dependent interactions 
acting between the quarks~\cite{stancu}. See a review
article~\cite{oka} for some theoretical attempts and problems 
which are discussed from a broad perspective. 

All the calculations performed so far in the constituent quark model 
were based on a variational principle which can be 
applied to a bound state problem. 
Without giving a link between such a bound state calculation 
and a decay width, it is hard to quantify the width 
of the $\Theta^+$. Though there are some attempts to predict 
the decay width of the $\Theta^+$ in the constituent quark 
model~\cite{enyo,carlson,hosaka}, either the unbound nature of 
the $\Theta^+$ is not taken into account or the assumption 
of its structure is too simple. 
Moreover, the correlated motion among the five quarks is usually 
truncated in the orbital, spin, isospin and color degrees 
of freedom or in some of them apart from a calculation~\cite{paris}.  
For example, a treatment of hidden color components 
is not quite clear. Though the colors of the 
quarks are uniquely coupled to a color singlet state in a baryon 
and a meson, we have several 
possibilities to get a color singlet pentaquark. Therefore it is 
desirable to carefully examine the role 
of the hidden color states in a 
variational calculation. This problem is avoided in 
Ref.~\cite{takeuchi} by averaging a confinement potential.

We will instead show that it is possible to 
predict the mass of the pentaquark even though the hidden color 
components are included in a calculation. 
A very recent quark model calculation with a scattering boundary 
condition~\cite{hiyama} is perhaps the only one that has undertaken 
a prediction of the width of the $\Theta^+$. 
It is claimed there that a sharp resonance with 
$\frac{1}{2}^{-}$ appears above 2 GeV, which is in disagreement with 
the other predictions~\cite{enyo,hosaka}.  In any case, the appearance of 
such a narrow resonance is hard to 
understand unless the $\Theta^+$ has negligibly small 
$NK$ component. It is fair to say that the quark model study has not so
far given a settled prediction for the decay width of the $\Theta^+$ though 
it seems to predict more or less a similar mass for it. 

The purpose of this study is to predict the mass and the 
decay width of the $\Theta^+$ in a few-body approach 
based on the quark model. We assume that the $\Theta^+$ is a 
five-particle system but not a heptaquark system like 
$K\pi N$~\cite{bicudo}, as discussed in \cite{estrada}.  
We calculate the mass and the width for some 
$J^P$ states assuming that the system has isospin 0 and is 
confined in a spatially small region. We pay a 
special attention to the unbound nature of the sought solution and 
perform a dynamical calculation which allows for not only 
spatial correlations among the quarks but also all possible 
spin-isospin-color configurations.  
The merit of employing the quark model is that it exploits 
the symmetry of 
the system and can test the validity of some models through the 
dynamical calculation.  
In this way, we can learn the extent to which the diquark 
model~\cite{jaffe} or the diquark-triquark model~\cite{karliner} 
is sound. Also our model can include the effect of $NK$ and $NK^*$ 
channels on the $\Theta^+$, which is vital to predict the width. 

We use a correlated Gaussian basis and a complete set of the 
spin-isospin-color configurations. To predict the mass and decay 
width of the $\Theta^+$, we use a real stabilization 
method~\cite{hazi} which utilizes square-integrable basis functions 
to localize the resonance. We use a stochastic variational 
method~\cite{svm,book} (SVM) to set up the basis set which describes 
resonance states. Our emphasis here is not to include continuum 
states explicitly but to localize the resonance in 
a simple way, which makes it possible to calculate the reduced width. 
We believe that to evaluate the decay width 
is still a worthwhile study, considering the present 
controversial situation as mentioned above. 

The composition of the paper is as follows.  
Our formalism is explained in sect.~\ref{model}. 
We use an OGE potential 
as the interaction between the quarks, as defined 
in sect.~\ref{hamiltonian}. 
Some details are given for the basis function in a particular representation 
in sect.~\ref{wavefunction} and for 
its transformation to other basis functions 
in sect.~\ref{trans.basis}. The application 
of the real stabilization method to the description of a resonance 
is presented in sect.~\ref{resonance.rsm}. Here we illustrate 
an emergence of a unique solution (resonance) in the bound-state-looking 
calculations. The results of calculation are presented in 
sect.~\ref{results}. The mass spectrum of the $\Theta^+$ is 
given in sect.~\ref{mass.spectrum}, the decay width is discussed 
in sect.~\ref{decay}, and the structure of the $\Theta^+$ is 
analyzed in sect.~\ref{structure}. A summary is given in sect.~\ref{summary}.

\section{Formalism}
\label{model}

\subsection{Hamiltonian}
\label{hamiltonian}
A Hamiltonian for the $\Theta^+$ reads as  
\begin{equation}
H=\sum_{i=1}^5m_i+\sum_{i=1}^5\frac{{\bfi p_i}^2}{2m_i}-T_{\mathrm{c.m.}}
+\sum_{i<j}V_{ij},
\label{quark.hamiltonian}
\end{equation}
where $m_i$ is the mass of the $i$th quark and $V_{ij}$ is the 
interaction potential between the quarks. 
The so-called natural units are used, 
so a length has a dimension of inverse energy 
(1 fm=$\frac{1}{197.3}$ MeV$^{-1}$). 
The kinetic energy of the quark is given in a nonrelativistic 
form and the kinetic energy of the center of mass motion, 
$T_{\mathrm{c.m.}}$,  
is subtracted from the Hamiltonian. The eigenvalue of the 
Hamiltonian corresponds to the intrinsic mass of the $\Theta^+$.   

The quark-quark potential used in this paper consists of a 
variant of OGE potential, 
a phenomenological confinement
potential and a zero-point energy term. 
It is taken from literatures: AL1 potential~\cite{al1}
\begin{equation}
V_{ij}=
-\frac{3}{8}(\lambda_i^{\rm C}\cdot\lambda_j^{\rm C})
\left(
-\frac{\kappa}{r_{ij}}+\lambda r_{ij}-\Lambda+\frac{2\pi\kappa'}{3m_im_j}
\frac{\exp(-\frac{r_{ij}^2}{\rho_{ij}^2})}{\pi^{3/2}\rho_{ij}^3}
\sigma_i\cdot\sigma_j
\right), 
\label{al1pot}
\end{equation}
with $\rho_{ij}=A\left(\frac{2m_im_j}{m_i+m_j}\right)^{-B}$ or TS potential~\cite{takeuchi}
\begin{eqnarray}
V_{ij} &= &
(\lambda_i^{\rm C}\cdot\lambda_j^{\rm C})\frac{\alpha_S}{4}
\left\{\frac{1}{r_{ij}}
-\frac{\mathrm{e}^{-\Lambda_g r_{ij}}}{r_{ij}}
-\left(\frac{\pi}{2m_i^2}+\frac{\pi}{2m_j^2}+\frac{2\pi}{3m_im_j}
(\sigma_i\cdot\sigma_j) \right)\frac{\Lambda_g^2}{4\pi}
\frac{\mathrm{e}^{-\Lambda_g r_{ij}}}{r_{ij}}
\right\}
\nonumber\\
&-&(\lambda_i^{\rm C}\cdot\lambda_j^{\rm C})
a_{\mathrm{conf}}r_{ij}+v_0.
\label{tspot}
\end{eqnarray}
Both potentials contain the color 
Coulomb and color magnetic terms. 

$\lambda_i^{\rm C}$ are the color SU(3) generators 
(Gell-Mann matrices) for the $i$th quark. 
In this paper an SU(3) representation is labeled by 
Elliott's convention~\cite{elliott} 
$\Gamma\!=\!(\lambda \mu)$ whose dimension is 
$d(\lambda \mu)\!=\!\frac{1}{2}(\lambda \!+\!1)(\mu \!+\!1)
(\lambda \!+\!\mu \!+\!2)$. 
A quark carries a color of (10) representation 
and an antiquark carries that of (01) representation. 
The color state of two quarks is either (20) (color symmetric) or 
(01) (color antisymmetric). Corresponding to these representations, 
the matrix of $(\lambda_i^{\rm C}\cdot\lambda_j^{\rm C})$ becomes 
\begin{equation}
(\lambda_i^{\rm C}\cdot\lambda_j^{\rm C})=
\left(
\begin{array}{cc}
\frac{4}{3} & 0  \\
0 & -\frac{8}{3} \\
\end{array}
\right).
\end{equation}
For the case of a quark-antiquark pair the color state is 
either (11) or (00), and the matrix 
of $(\lambda_i^{\rm C}\cdot\lambda_j^{\rm C})$ corresponding to 
these states is 
\begin{equation}
(\lambda_i^{\rm C}\cdot\lambda_j^{\rm C})=
\left(
\begin{array}{cc}
\frac{2}{3} & 0  \\
0 & -\frac{16}{3} \\
\end{array}
\right).
\end{equation}
In cases where the two quarks are in the color state (20) 
or the quark-antiquark pair is in (11), 
$(\lambda_i^{\rm C}\cdot\lambda_j^{\rm C})$ gives 
a sign opposite to the physical cases which appear 
in a baryon and a meson, so those color states play 
a role of deconfining the quarks at large distances. 
We will not exclude such color states, however, from 
the beginning as 
they may be important in the $\Theta^+$ comprising 
four quarks and one antiquark in a spatially confined region.  

For a color singlet system, the zero-point energy term has an 
expectation value 
\begin{equation}
\left\langle \sum_{i<j}(\lambda_i^{\rm C}\cdot\lambda_j^{\rm
 C})\right\rangle_{(00)000}
=\left\langle \frac{1}{2}
\left(\sum_i\lambda_i^{\rm C}\right)^2-\frac{1}{2}\sum_i 
(\lambda_i^{\rm C})^2\right\rangle_{(00)000}=-\frac{8}{3}n,
\end{equation}
where the subgroup label of the color singlet state is denoted 000, 
and where $n$ is the total number of quarks and antiquarks 
contained in the system and it determines the expectation value 
independently of the way of constructing the color state. 
Thus the zero-point energies of the AL1 potential are 
$-2\Lambda$, $-3\Lambda$
and $-5\Lambda$, for mesons, baryons and the $\Theta^+$,
respectively. In the case of the TS potential, 
$v_0$ is varied depending on $n$: $v_0=V_0'$ for the meson, 
$v_0=3V_0$ for the baryon, and $v_0=5V_0$ for the $\Theta^+$. 
To change the parameter of the zero-point energy results in 
simply shifting the mass but never alters the
eigenfunction of the Hamiltonian. This property will be used 
in sect.~\ref{decay} to estimate the decay width of the $\Theta^+$.

We list in Table~\ref{tab.baryon-meson} the masses of 
some mesons and baryons calculated using the 
SVM~\cite{svm,book}. The parameters of the potential 
are also given in the table. It is found 
that both potentials give the results which agree 
reasonably well with the observed masses of $K, K^*, N(939)$ 
and $\Delta(1232)$. The $N\!+\!K$ threshold is 
1486 and 1451 MeV for the AL1 and TS potentials, respectively. 
The masses of the strange baryons 
predicted with the TS potential are considerably large compared 
to the experiment, which is 
probably due to the fact that the mass of the strange quark is 
taken to be large. The masses of some baryons such as  
$N(1440)$ and $\Lambda(1405)$ are found 
to be too large. A flavor-dependent potential or some multi-quark 
configurations may be needed to reproduce these
masses~\cite{glozman96,glozman98}. The root mean square (rms) 
radius of 
the quark distribution is also calculated and listed in 
the table. The calculated size of $N(939)$ is apparently 
too small, so we must make allowance for this underestimation in 
setting a channel radius which is needed to calculate the decay 
width of the $\Theta^+$.

\begin{table}[b]
\caption{The masses $M$ and rms radii $\sqrt{\langle r^2\rangle}$ 
of mesons and baryons. $L$ is the total orbital angular momentum assumed 
in the calculation. The parameters for the AL1 potential~\cite{al1} 
are $m_{ud}\!=\!315$ MeV, $m_s\!=\!577$ MeV, $\kappa\!=\!0.5069$, 
$\lambda\!=\!0.1653$ GeV$^2$, $\Lambda\!=\!0.8321$ GeV, 
$\kappa'\!=\!1.8609$, 
$B\!=\!0.2204$, and $A\!=\!1.6553$ GeV$^{B-1}$, while those for the 
TS potential~\cite{takeuchi} are $m_{ud}\!=\!313$ MeV, $m_s\!=\!680$ MeV, 
$\alpha_S\!=\!1.72$, 
$\Lambda_g\!=\!3$ fm$^{-1}$, $\alpha_{\rm conf}\!=\!172.4$ MeVfm$^{-1}$, 
$V_0\!=\!-345.5$ MeV, and $V_0'\!=\!-742.7$ MeV.} 
\begin{center}
\begin{tabular}{cccccccccc}
\hline\hline
 & & & & \multicolumn{2}{c}{AL1} & & & \multicolumn{2}{c}{TS}\\
\cline{5-6}\cline{9-10}
Particle & & $L$ & & $M$ [MeV] & $\sqrt{\langle r^2\rangle}$ [fm]
& & & $M$ [MeV] & $\sqrt{\langle r^2\rangle}$ [fm]\\
\hline
$\pi$(138)    & & 0 & & 138  & 0.30 & & & 106  & 0.32 \\
$\rho$(770)   & & 0 & & 769  & 0.46 & & & 660  & 0.45 \\
$K$(496)      & & 0 & & 491  & 0.31 & & & 514  & 0.33 \\
$K^*$(892)    & & 0 & & 903  & 0.42 & & & 813  & 0.41 \\
$N$(939)      & & 0 & & 995  & 0.49 & & & 937  & 0.50 \\
$N$(1440)     & & 0 & & 1722 & 0.75 & & & 1755 & 0.73 \\
$\Delta$(1232)& & 0 & & 1307 & 0.58 & & & 1229 & 0.56 \\
$\Lambda$(1116)& &0 & & 1148 & 0.47 & & & 1266 & 0.49 \\
$\Lambda$(1405)& &1 & & 1522 & 0.58 & & & 1736 & 0.60 \\
$\Sigma$(1193)& & 0 & & 1229 & 0.49 & & & 1381 & 0.51 \\
$\Xi$(1318)   & & 0 & & 1349 & 0.44 & & & 1659 & 0.49 \\
$\Omega$(1672)& & 0 & & 1675 & 0.48 & & & 2084 & 0.50 \\
\hline\hline
\end{tabular}
\end{center}
\label{tab.baryon-meson}
\end{table}

\subsection{Basis function}
\label{wavefunction}
Our Hamiltonian commutes with the total orbital angular 
momentum, the total spin and the total isospin. 
Thus the $\Theta^+$ state is specified with their quantum numbers  
$L, S$ and $T$.  By letting $J$ and $P$ denote the total angular 
momentum and parity of the $\Theta^+$, we may write its wave 
function as an antisymmetrized product of the orbital, spin, 
isospin and color parts:
\begin{equation}
\Psi_{JM}^P={\cal A}\Bigg\{[\psi_L^{\rm (orbital)}\psi_S^{\rm (spin)}]_{JM}\psi_{TM_T}^{\rm (isospin)}
\psi_{(00)000}^{\rm (color)}\Bigg\},
\label{totalwf}
\end{equation}
where ${\cal A}$ is an antisymmetrizer of the four quarks ($uudd$) 
labeled 1, 2, 3, and 4 ($\bar{s}$ is labeled 5) normalized to 
${\cal A}^2={\cal A}$,  
the square bracket 
$[\psi_L^{\rm (orbital)}\psi_S^{\rm (spin)}]_{JM}$ 
stands for the angular momentum coupling. 
The $\Theta^+$ masses with different $J$ values for a given set 
of $L$ and $S$ values are degenerate in our model. 

Now we specify each part of Eq.~(\ref{totalwf}) in detail. 
It is vital to allow for various 
types of correlation among the quarks. For the orbital 
part, we use an 
explicitly correlated Gaussian basis in a global 
representation~\cite{svm,book,global}:
\begin{eqnarray}
\psi_{LM_L}^{\rm (orbital)} \sim \phi_{LM_L}(A,u,{\bfi x})
&=&{\exp}\left\{-\frac{1}{2}\widetilde{{\bfi x}}A{\bfi x}\right\}
{\cal Y}_{LM_L}(\widetilde{u}{\bfi x})
\nonumber \\
&\equiv&{\exp}\left\{-\frac{1}{2}\sum_{i,j=1}^4A_{ij}{\bfi
	       x}_i\cdot{\bfi x}_j\right\}{\cal
Y}_{LM_L}\left(\sum_{i=1}^4 u_i{\bfi x}_i\right).
\label{cgwf}
\end{eqnarray}
where ${\bfi x}\!=\!({\bfi x}_1, {\bfi x}_2, {\bfi x}_3, {\bfi x}_4)$ stands
for a set of the intrinsic Jacobi 
coordinates other than the center of mass coordinate of the system 
and it is defined from the single-particle coordinates of the 
quarks, $({\bfi r}_1, {\bfi r}_2, \ldots, {\bfi r}_5)$, as usual. 
The angular motion of the system is described with a solid 
spherical harmonics ${\cal Y}_{\ell m}({\bfi r})=r^{\ell}Y_{\ell m}
(\hat{\bfi r})$. 
Here a positive-definite, symmetric matrix $A$ and a vector $u$ 
are parameters which define the shape of the orbital part of the 
basis wave function. Note that $u$ 
serves to define an appropriate coordinate responsible for 
the rotational motion of the system and 
that the spherical part of the 
orbital function can equivalently be expressed in terms of the 
relative distance vectors of the quarks: 
\begin{equation}
{\exp}\left\{-\frac{1}{2}\widetilde{{\bfi x}}A{\bfi x}\right\}=
{\exp}\, \Big\{-\sum_{l>k=1}^5\alpha_{kl}({\bfi r}_k-{\bfi r}_l)^2\Big\}
\end{equation}
with appropriate $\alpha_{kl}$'s. We use the correlated Gaussian 
because there are many examples which demonstrate its power for an 
accurate description of few-particle systems~\cite{svm,book}. 
Clearly our orbital part is translation-invariant, so our 
theory is free from any spurious center of mass motion. 
Note that the parity of the $\Theta^+$ 
is given by $P=(-1)^{L+1}$ in the present formalism. 

One of the most natural coupling schemes for the spin, isospin and 
color parts is to follow a successive coupling. The spin part 
may be expressed as 
\begin{equation}
\psi_{SM_S}^{\rm (spin)}\sim \chi^{\rm SC}_{(S_{12}S_{123}S_{1234})SM_S} = 
\Bigg[\Big[\big[[\chi_{\frac{1}{2}}(1)\chi_{\frac{1}{2}}(2)]_{S_{12}}\chi_{\frac{1}{2}}(3)\big]_{S_{123}}
\chi_{\frac{1}{2}}(4)\Big]_{S_{1234}}\chi_{\frac{1}{2}}(5)\Bigg]_{SM_S},
\end{equation} 
where $\chi_{\frac{1}{2}m_s}(i)$ is the spin function of the 
$i$th quark. Table~\ref{tab.spin} 
lists possible sets of $(S_{12}, S_{123}, S_{1234})$ for $S\!=\!\frac{1}{2}$ and $\frac{3}{2}$.  
For the isospin part we assume $T\!=\!0,\, M_T\!=\!0$. 
Because the $\bar{s}$ does not carry an 
isospin, the isospin part can be given by coupling four 
$\frac{1}{2}$ angular momenta:
\begin{equation}
\psi_{TM_T}^{\rm (isospin)}\sim \xi^{\rm SC}_{(T_{12}T_{123}0)T=0M_T=0} = 
\Big[\big[[\xi_{\frac{1}{2}}(1)\xi_{\frac{1}{2}}(2)]_{T_{12}}\xi_{\frac{1}{2}}(3)\big]_{T_{123}}
\xi_{\frac{1}{2}}(4)\Big]_{00}\xi(5),
\end{equation} 
where $\xi_{\frac{1}{2}m_t}(i)$ is the isospin function of the 
$i$th quark and $\xi(5)$ 
the ``isospin" function of the $\bar{s}$. 
Possible values of $(T_{12}, T_{123})$ that enable one to 
make $T\!=\!0,\, M_T\!=\!0$ are $(0, \frac{1}{2})$ and 
$(1, \frac{1}{2})$. Similarly, the color 
part can be given, in the successive coupling, as 
\begin{eqnarray}
\psi_{(00)000}^{\rm (color)} &\sim&
 C^{\rm SC}_{(\Gamma_{12}\Gamma_{123}\Gamma_{1234})(00)000} \nonumber \\
&=& 
\Bigg[\Big[\big[[C_{(10)}(1) C_{(10)}(2)]_{\Gamma_{12}}
C_{(10)}(3)\big]_{\Gamma_{123}}
C_{(10)}(4)\Big]_{\Gamma_{1234}} C_{(01)}(5)\Bigg]_{(00)000},
\end{eqnarray}
where $C(i)$ stands for the color function of the $i$th quark, and 
the square bracket $[C_{\Gamma_1} C_{\Gamma_2}]_{\Gamma}$ 
denotes the SU(3) coupling of two functions with SU(3) 
irreducible 
representations $\Gamma_1$ and $\Gamma_2$ to that of a definite 
SU(3) representation $\Gamma$. Here $\Gamma_{1234}$ must be 
(10) to make a color singlet $\Theta^+$. Possible sets of 
$(\Gamma_{12},\Gamma_{123},\Gamma_{1234}=(10))$ 
are listed in Table~\ref{tab.color}. As seen from the table, 
three channels make a complete set for the color space of 
the $\Theta^+$. 

\begin{table}[t]
\caption{Intermediate spin labels in different models}
\label{tab.spin}
\begin{center}
\begin{tabular}{lccccccccccccc}
\hline\hline
$S=\frac{1}{2}$ & & & & & & & & & & & & & \\
\hline
 & \multicolumn{3}{c}{Successive coupling} & & & 
 \multicolumn{3}{c}{Diquark-diquark type} & & & \multicolumn{3}{c}{Baryon-Meson type} \\
\cline{2-4}\cline{7-9}\cline{12-14}
 & $S_{12}\ \ $ & $S_{123}\ $  & $S_{1234}$ & & & $S_{12}\ \ $  & $S_{34}\
 $ &
 $S_{1234}$ & & & $S_{12}\ \ $ & $S_{123}\ $ & $S_{45}$ \\ 
\cline{2-14}
 & 0 & $\frac{1}{2}$  & 0 & & & 0  & 0 & 0 & & & 0 & $\frac{1}{2}$ & 0 \\
 & 0 & $\frac{1}{2}$  & 1 & & & 0  & 1 & 1 & & & 0 & $\frac{1}{2}$ & 1 \\
 & 1 & $\frac{1}{2}$  & 0 & & & 1  & 1 & 0 & & & 1 & $\frac{1}{2}$ & 0 \\
 & 1 & $\frac{1}{2}$  & 1 & & & 1  & 0 & 1 & & & 1 & $\frac{1}{2}$ & 1 \\
 & 1 & $\frac{3}{2}$  & 1 & & & 1  & 1 & 1 & & & 1 & $\frac{3}{2}$ & 1 \\
\hline\hline
$S=\frac{3}{2}$ & & & & & & & & & & & & & \\
\hline
 & \multicolumn{3}{c}{Successive coupling} & & &
 \multicolumn{3}{c}{Diquark-diquark type} & & & \multicolumn{3}{c}{Baryon-Meson type} \\
\cline{2-4}\cline{7-9}\cline{12-14}
 & $S_{12}\ \ $ & $S_{123}\ $  & $S_{1234}$ & & & $S_{12}\ \ $  & $S_{34}\
 $ &
 $S_{1234}$ & & & $S_{12}\ \ $ & $S_{123}\ $ & $S_{45}$ \\ 
\cline{2-14}
 & 1 & $\frac{3}{2}$  & 2 & & & 1  & 1 & 2 & & & 1 & $\frac{3}{2}$ & 0 \\
 & 1 & $\frac{3}{2}$  & 1 & & & 1  & 0 & 1 & & & 1 & $\frac{3}{2}$ & 1 \\
 & 0 & $\frac{1}{2}$  & 1 & & & 0  & 1 & 1 & & & 0 & $\frac{1}{2}$ & 1 \\
 & 1 & $\frac{1}{2}$  & 1 & & & 1  & 1 & 1 & & & 1 & $\frac{1}{2}$ & 1 \\
\hline\hline
\end{tabular}
\end{center}
\end{table}

To sum up, we may express a trial wave function for the 
$\Theta^+$ as 
\begin{equation}
\Psi_{JM}^P=\sum_{i=1}^{\cal K} C_i\Phi_i,
\end{equation}
with the basis function 
\begin{equation}
\Phi_i={\cal A}\Bigg\{
[{\phi}_{L}(A,u,{\bfi x})\chi_{(S_{12}S_{123}S_{1234})S}]_{JM}\xi_{(T_{12}T_{123}0)00}C_{(\Gamma_{12}\Gamma_{123}\Gamma_{1234})(00)000}\Bigg\},
\label{basisfunction}
\end{equation}
where $i$ stands for a set of $(A, u, S_{12}, S_{123}, S_{1234}, T_{12}, \Gamma_{12}, 
\Gamma_{123})$. Here $T_{123}$ and $\Gamma_{1234}$ are omitted as 
they have to be $\frac{1}{2}$ and $(10)$, respectively. Note that 
the 4$\times 4$ matrix $A$ is specified by 
ten parameters $(A_{11}, A_{12}, \ldots, A_{44})$ or equivalently 
$(\alpha_{12}, \alpha_{13}, \ldots, \alpha_{45})$ 
and the $u$ by three parameters (as it can be set to a unit 
vector, $\widetilde{u}u\!=\!1$, without loss of generality). 
Thus we have 13 (10) continuous parameters for 
the basis function with $L\! \ne \!0$ ($L\!=\!0$) and in 
addition the spin-isospin-color channel label which is one 
of $5\!\times\! 2\times\! 3=30$ channels 
for $S\!=\!\frac{1}{2}$ or $4\!\times \!2\times \!3=24$ channels 
for $S\!=\!\frac{3}{2}$.  
We stress that all the possibilities satisfying the prescribed 
quantum numbers ($S,\, T\!=\!0,\, \Gamma\!=\!(00)$) can be 
taken into account 
in the present formalism. Once the basis functions 
$\Phi_i$ are set up, the coefficients $C_i$ and the eigenvalue 
$M$ of the Hamiltonian 
can be determined by solving the following eigenvalue problem:
\begin{equation}
\sum_{j=1}^{\cal K}[\langle \Phi_i|H|\Phi_j\rangle - 
M\langle \Phi_i|\Phi_j\rangle] \, C_j=0.
\end{equation}

The label SC is omitted from the spin-isospin-color parts 
in Eq.~(\ref{basisfunction}) because other coupling schemes may be used 
equally well as explained below. 

\begin{table}[t]
\caption{Intermediate color SU(3) labels in different models}
\label{tab.color}
\begin{center}
\begin{tabular}{ccccccccccccc}
\hline\hline
 \multicolumn{3}{c}{Successive coupling} & & &
 \multicolumn{3}{c}{Diquark-diquark type} & & & \multicolumn{3}{c}{Baryon-Meson type} \\
\cline{1-3}\cline{6-8}\cline{11-13}
 $\Gamma_{12}\ \ $ & $\Gamma_{123}\ $  & $\Gamma_{1234}$ & & &
 $\Gamma_{12}\ \ $  & $\Gamma_{34}\ $ &
 $\Gamma_{1234}$ & & & $\Gamma_{12}\ \ $ & $\Gamma_{123}\ $ & $\Gamma_{45}$ \\ 
\cline{1-13}
  (01) & (00)  & (10) & & & (01)  & (01) & (10) & & & (01) & (00) & (00) \\
  (01) & (11)  & (10) & & & (01)  & (20) & (10) & & & (01) & (11) & (11) \\
  (20) & (11)  & (10) & & & (20)  & (01) & (10) & & & (20) & (11) & (11) \\
\hline\hline
\end{tabular}
\end{center}
\end{table}

\subsection{Transformation of basis }
\label{trans.basis}

The successive coupling scheme introduced in sect.~\ref{wavefunction} is 
systematic in its construction. The basis functions in that scheme 
constitute a complete set 
for specified quantum numbers and make it possible 
to represent any coupling schemes of physical interest. 

The $\Theta^+$ is considered a system of 
two diquarks and the $\bar{s}$ in the diquark model~\cite{jaffe}, 
so it is useful to define 
the following coupling scheme, e.g., for the spin part 

\begin{equation}
\chi^{\rm DD}_{(S_{12}S_{34}S_{1234})SM_S} = 
\Big[\big[[\chi_{\frac{1}{2}}(1)\chi_{\frac{1}{2}}(2)]_{S_{12}}[\chi_{\frac{1}{2}}(3)\chi_{\frac{1}{2}}(4)]_{S_{34}}\big]_{S_{1234}}\chi_{\frac{1}{2}}(5)\Big]_{SM_S}.
\end{equation} 
Here one diquark has spin $S_{12}$ and the other 
diquark $S_{34}$, and they are coupled to $S_{1234}$. 
We list possible sets of $(S_{12}, S_{34}, S_{1234})$ in 
Table~\ref{tab.spin}. 
As mentioned, this spin function can be expressed in terms of 
combinations of the spin functions in the successive coupling. 
This expansion can be represented as follows  
(by replacing the spin label $S$ with $J$): 
\begin{eqnarray}
\begin{minipage}{12.5cm}
\begin{center}
\begin{picture}(400,100)(0,100)
\put(20,120){\line(-1,4){8}}
\put(12,152){\line(3,4){19}}
\put(20,120){\line(1,5){11.5}}
\put(31.4,177.4){\line(5,1){31}}
\put(31,176.4){\line(4,-1){52}}
\put(62,184){\line(1,-1){22}}
\put(20,120){\line(3,2){64}}
\put(84,162){\line(1,-6){4.7}}
\put(20,120){\line(5,1){68.2}}
\put(182,141){\makebox(10,10)
{$\displaystyle =\sum_{J_{123}}\,U(J_{12}J_3J_{1234}J_4;J_{123}J_{34})$}}
\put(285,120){\line(-1,4){8}}
\put(277,152){\line(3,4){19}}
\put(285,120){\line(1,5){11.5}}
\put(296.4,177.4){\line(5,1){31}}
\put(285,120){\line(2,3){42}}
\put(327,184){\line(1,-1){22}}
\put(285,120){\line(3,2){64}}
\put(349,162){\line(1,-6){4.7}}
\put(285,120){\line(5,1){68.2}}
\put(5,130){\makebox(10,10){$J_1$}}
\put(11,165){\makebox(10,10){$J_2$}}
\put(40,183){\makebox(10,10){$J_3$}}
\put(74,175){\makebox(10,10){$J_4$}}
\put(90,147){\makebox(10,10){$J_5$}}
\put(57,114){\makebox(10,10){$JM$}}
\put(32,146){\makebox(10,10){$J_{12}$}}
\put(51,156){\makebox(10,10){$J_{34}$}}
\put(62,135){\makebox(10,10){$J_{1234}$}}
\put(270,130){\makebox(10,10){$J_1$}}
\put(276,165){\makebox(10,10){$J_2$}}
\put(305,183){\makebox(10,10){$J_3$}}
\put(339,175){\makebox(10,10){$J_4$}}
\put(355,147){\makebox(10,10){$J_5$}}
\put(322,114){\makebox(10,10){$JM$}}
\put(297,160){\makebox(10,10){$J_{12}$}}
\put(321,160){\makebox(10,10){$J_{123}$}}
\put(327,135){\makebox(10,10){$J_{1234}$}}
\end{picture}
\end{center}
\end{minipage}
\label{diquark-successive}
\end{eqnarray}
Here a Racah coefficient in unitary form, $U$, is expressed 
in terms of the 6$j$ symbol
\begin{eqnarray}
U(J_1J_2 J J_3;J_{12} J_{23})&=&
\langle (J_1J_2)J_{12}, J_3; JM|J_1,(J_2J_3)J_{23}; JM \rangle
\nonumber\\
&=&(-1)^{J_1+J_2+J+J_3}\sqrt{(2J_{12}+1)(2J_{23}+1)}
\left\{
\begin{array}{lll}
 J_1 & J_2 & J_{12} \\
 J_3 & J   & J_{23} \\
\end{array}
\right\}.
\end{eqnarray}
To be more explicit, the spin function in the diquark model 
is transformed to the one in the successive coupling scheme by 
the matrix 
\begin{equation}
\left(
\begin{array}{ccccc}
1 & 0 & 0 & 0 & 0  \\
0 & 1 & 0 & 0 & 0  \\
0 & 0 & 1 & 0 & 0  \\
0 & 0 & 0 & -\sqrt{\frac{1}{3}} & \sqrt{\frac{2}{3}}  \\
0 & 0 & 0 & \sqrt{\frac{2}{3}} & \sqrt{\frac{1}{3}}  \\
\end{array}
\right)\ \ \ \ \ \  {\rm for}\ S=\frac{1}{2},
\end{equation}
and 
\begin{equation}
\left(
\begin{array}{cccc}
1 & 0 & 0 & 0   \\
0 & \sqrt{\frac{2}{3}} & 0 & -\sqrt{\frac{1}{3}}  \\
0 & 0 & 1 & 0   \\
0 & \sqrt{\frac{1}{3}} & 0 & \sqrt{\frac{2}{3}}   \\
\end{array}
\right)\ \ \ \ \ \  {\rm for}\ S=\frac{3}{2}.
\end{equation}
Here the spin functions in both models are arranged in order of 
those defined in Table~\ref{tab.spin}.  

Similarly, the color function in the diquark model 
can be expressed in terms of the successive coupling scheme. 
Possible sets of $(\Gamma_{12}, \Gamma_{34}, \Gamma_{1234})$ 
in the diquark model are listed in Table~\ref{tab.color}. 
The transformation between the diquark coupling and the 
successive coupling can be performed using the 
formula~(\ref{diquark-successive}) (by replacing the Racah 
coefficient with the corresponding one in the SU(3) 
algebra~\cite{hecht,draayer}). Here the angular momentum label $J$ 
should be understood to denote the color SU(3) label 
$\Gamma$. Note that all the SU(3) couplings appearing here 
are multiplicity-free. 
The color function in the diquark model is expressed in terms of 
the one in the successive coupling through the matrix  
\begin{equation}
\left(
\begin{array}{ccc}
-\sqrt{\frac{1}{3}} & \sqrt{\frac{2}{3}} & 0   \\
\sqrt{\frac{2}{3}} & \sqrt{\frac{1}{3}} & 0   \\
0 & 0 & 1   \\
\end{array}
\right).
\end{equation}

A baryon-meson ($q^3$-$q{\bar q}$) coupling scheme is also important. 
Here three among the four quarks are coupled to a state with a 
definite spin, isospin and 
color, and the remaining quark and the $\bar{s}$ are coupled to 
a state with given quantum numbers, and finally they are 
coupled to a resultant state. Again we can show 
this coupling scheme in a pictorial way and express it 
in terms of the successive coupling: 

\begin{eqnarray}
\begin{minipage}{12.5cm}
\begin{center}
\begin{picture}(400,100)(0,100)
\put(20,120){\line(-1,4){8}}
\put(12,152){\line(3,4){19}}
\put(20,120){\line(1,5){11.5}}
\put(31.4,177.4){\line(5,1){31}}
\put(20,120){\line(2,3){42}}
\put(62,184){\line(1,-1){22}}
\put(62.5,183){\line(1,-2){24.5}}
\put(84,162){\line(1,-6){4.7}}
\put(20,120){\line(5,1){68.2}}
\put(178,144){\makebox(10,10)
{$\displaystyle=\sum_{J_{1234}}\, U(J_{123}J_4JJ_5;J_{1234}J_{45})$}}
\put(275,120){\line(-1,4){8}}
\put(267,152){\line(3,4){19}}
\put(275,120){\line(1,5){11.5}}
\put(286.4,177.4){\line(5,1){31}}
\put(275,120){\line(2,3){42}}
\put(317,184){\line(1,-1){22}}
\put(275,120){\line(3,2){64}}
\put(339,162){\line(1,-6){4.7}}
\put(275,120){\line(5,1){68.2}}
\put(5,130){\makebox(10,10){$J_1$}}
\put(11,165){\makebox(10,10){$J_2$}}
\put(40,183){\makebox(10,10){$J_3$}}
\put(74,175){\makebox(10,10){$J_4$}}
\put(90,147){\makebox(10,10){$J_5$}}
\put(57,114){\makebox(10,10){$JM$}}
\put(32,160){\makebox(10,10){$J_{12}$}}
\put(45,140){\makebox(10,10){$J_{123}$}}
\put(63,149){\makebox(10,10){$J_{45}$}}
\put(260,130){\makebox(10,10){$J_1$}}
\put(266,165){\makebox(10,10){$J_2$}}
\put(295,183){\makebox(10,10){$J_3$}}
\put(329,175){\makebox(10,10){$J_4$}}
\put(345,147){\makebox(10,10){$J_5$}}
\put(312,114){\makebox(10,10){$JM$}}
\put(287,160){\makebox(10,10){$J_{12}$}}
\put(311,160){\makebox(10,10){$J_{123}$}}
\put(317,135){\makebox(10,10){$J_{1234}$}}
\end{picture}
\end{center}
\end{minipage}
\end{eqnarray}
Possible sets of $(J_{12}, J_{123}, J_{45})$ values 
in the baryon-meson coupling are listed 
in Table~\ref{tab.spin} for the spin part ($\chi^{\rm BM}$) 
and in Table~\ref{tab.color} 
for the color part ($C^{\rm BM}$), respectively. Note that $\Gamma_{123}$ and 
$\Gamma_{45}$ are not necessarily (00), that is, a colored baryon 
and a colored meson must be included to obtain a complete basis 
for a color singlet pentaquark. 
The spin function in the baryon-meson model is transformed to 
the one in the successive coupling by   
\begin{equation}
\left(
\begin{array}{ccccc}
-\frac{1}{2} & \frac{\sqrt{3}}{2} & 0 & 0 & 0  \\
\frac{\sqrt{3}}{2} & \frac{1}{2} & 0 & 0 & 0  \\
0 & 0 & -\frac{1}{2} & \frac{\sqrt{3}}{2} & 0  \\
0 & 0 & \frac{\sqrt{3}}{2} & \frac{1}{2} & 0  \\
0 & 0 & 0 & 0 & 1  \\
\end{array}
\right)\ \ \ \ \ \  {\rm for}\ S=\frac{1}{2},
\end{equation}
and 
\begin{equation}
\left(
\begin{array}{cccc}
\sqrt{\frac{5}{8}} & -\sqrt{\frac{3}{8}} & 0 & 0  \\
\sqrt{\frac{3}{8}} & \sqrt{\frac{5}{8}} & 0 & 0   \\
0 & 0 & 1 &  0  \\
0 & 0 & 0 & 1  \\
\end{array}
\right)\ \ \ \ \ \  {\rm for}\ S=\frac{3}{2}.
\end{equation}
The color function in the baryon-meson model is transformed 
to that of the successive coupling by a unit matrix.  

The Karliner-Lipkin model proposed in~\cite{karliner} 
considers the $\Theta^+$ as 
a coupled system of a $ud$ diquark and a $ud\bar s$ triquark. Here 
the $ud$ pair in the diquark has $S_{12}\!=\!0,\, T_{12}\!=\!0, \, 
\Gamma_{12}\!=\!(01)$, and the other pair in the triquark has 
$S_{34}\!=\!1,\, T_{34}\!=\!0, \, \Gamma_{34}\!=\!(20)$. The 
quantum numbers of the triquark is $S_{345}\!=\!\frac{1}{2},\, 
T_{345}\!=\!0, \, \Gamma_{345}\!=\!(10)$. Recoupling this 
configuration to the diquark-diquark model, we find that the 
Karliner-Lipkin model is 
reduced to one particular basis in the diquark model with 
$S_{1234}\!=\!1,\, T_{1234}\!=\!0, \, \Gamma_{1234}\!=\!(10)$, namely  
\begin{eqnarray}
\begin{minipage}{12.5cm}
\begin{center}
\begin{picture}(400,100)(0,100)
\put(20,120){\line(-1,4){8}}
\put(12,152){\line(3,4){19}}
\put(20,120){\line(1,5){11.5}}
\put(31.4,177.4){\line(5,1){31}}
\put(31,176.4){\line(4,-1){52}}
\put(62,184){\line(1,-1){22}}
\put(31.2,177){\line(4,-3){57}}
\put(84,162){\line(1,-6){4.7}}
\put(20,120){\line(5,1){68.2}}
\put(123,141){\makebox(1,1){$=$}}
\put(150,120){\line(-1,4){8}}
\put(142,152){\line(3,4){19}}
\put(150,120){\line(1,5){11.5}}
\put(161.4,177.4){\line(5,1){31}}
\put(161,176.4){\line(4,-1){52}}
\put(192,184){\line(1,-1){22}}
\put(150,120){\line(3,2){64}}
\put(214,162){\line(1,-6){4.7}}
\put(150,120){\line(5,1){68.2}}
\put(57,114){\makebox(5,5){{\scriptsize $\frac{1}{2}0(00)$}}}
\put(33,152){\makebox(5,5){{\scriptsize $00(01)$}}}
\put(54,170){\makebox(5,5){{\scriptsize $10(20)$}}}
\put(97,149){\makebox(5,5){{\scriptsize $\frac{1}{2}\_(01)$}}}
\put(67,144){\makebox(5,5){{\scriptsize $\frac{1}{2}0(10)$}}}
\put(162,152){\makebox(5,5){{\scriptsize $00(01)$}}}
\put(185,169){\makebox(5,5){{\scriptsize $10(20)$}}}
\put(228,149){\makebox(5,5){{\scriptsize $\frac{1}{2}\_(01)$}}}
\put(188,114){\makebox(5,5){{\scriptsize $\frac{1}{2}0(00)$}}}
\put(183,140){\makebox(5,5){{\scriptsize $10(10)$}}}
\end{picture}
\end{center}
\end{minipage}
\label{klmodel}
\end{eqnarray}

The calculation of the decay width of the $\Theta^+$ will be 
reduced to an overlap of the $\Theta^+$ wave function with 
an $nK^+$ channel wave function. See sect.~\ref{decay}. 
Here we write the spin-isospin-color part of the $nK^+$ channel 
wave function in terms of the successive coupling scheme. Both 
$n$ and $K^+$ have isospin $\frac{1}{2}$ and only the 
isospin-singlet 
component of the coupled state contributes to the overlap with the 
$\Theta^+$ wave function. 
The nucleon ($N$) is symmetric in its orbital space 
as well as in the spin-isospin space. 
Combining the transformation matrices given above, we can 
express the spin-isospin-color part of   
$n$ and $K^+$, $\psi(n)\psi(K^+)$, in terms of 
the successive coupling as follows (by omitting the isospin-triplet 
component): 
\begin{eqnarray} 
[\psi(n)\psi(K^+)]_{\frac{1}{2}m} &\to& 
-\frac{1}{\sqrt{2}}\frac{1}{\sqrt{2}}
\Bigg\{\Big(-\frac{1}{2}\chi^{\rm SC}_{(0\frac{1}{2}0)\frac{1}{2}m}+
\frac{\sqrt{3}}{2}\chi^{\rm SC}_{(0\frac{1}{2}1)\frac{1}{2}m}\Big)
\xi^{\rm SC}_{(0\frac{1}{2}0)00}
\nonumber \\
&+&
\Big(-\frac{1}{2}\chi^{\rm SC}_{(1\frac{1}{2}0)\frac{1}{2}m}+
\frac{\sqrt{3}}{2}\chi^{\rm SC}_{(1\frac{1}{2}1)\frac{1}{2}m}\Big)
\xi^{\rm SC}_{(1\frac{1}{2}0)00}
\Bigg\}C^{\rm SC}_{((01)(00)(10))(00)000}.
\end{eqnarray}
Here the first $-\frac{1}{\sqrt{2}}$ factor comes from the 
Clebsch-Gordan coefficient of the isospin-singlet coupling and the 
second $\frac{1}{\sqrt{2}}$ factor from the symmetric 
spin-isospin function of the nucleon. 

We briefly comment on the 
transformation of the orbital part. The coordinate  
appropriate for the diquark model is defined as 
\begin{eqnarray}
&&{\bfi y}_1={\bfi r}_1-{\bfi r}_2,\ \ \ \ {\bfi y}_2={\bfi r}_3-{\bfi
 r}_4,\ \ \ \ {\bfi y}_3=\frac{1}{2}({\bfi r}_1+{\bfi r}_2)-\frac{1}{2}({\bfi
 r}_3+{\bfi r}_4),
\nonumber \\
&& {\bfi y}_4=\frac{1}{4}({\bfi r}_1+{\bfi r}_2
+{\bfi r}_3+{\bfi r}_4)-{\bfi r}_5.
\end{eqnarray}
An orbital function in the diquark model may be expressed as
$\phi_{LM_L}(A',u',{\bfi y})$.  
Because ${\bfi y}$ is related to ${\bfi x}$ by a linear 
transformation, ${\bfi y}=T{\bfi x}$, with an appropriate 
$4\times 4$ matrix $T$, 
this orbital function can be reduced to the one in the 
${\bfi x}$ representation: 
\begin{equation}
\phi_{LM_L}(A',u',{\bfi y})=\phi_{LM_L}(A'_T,u'_T,{\bfi x}),
\end{equation}
with
\begin{equation}
A'_T=\widetilde{T}A'T,\ \ \ \ \  u'_T=\widetilde{T}u'.
\end{equation}
Here $\widetilde{T}$ is a transposed matrix of $T$. 
Therefore, we do not need to introduce a new diquark orbital function
explicitly but can take into account it by simply adopting
$A'_T$ and $u'_T$ as $A$ and $u$ in Eq.~(\ref{cgwf}).  
It is clear that an orbital function appropriate for the 
baryon-meson model can be expressed in terms of the 
correlated Gaussian~(\ref{cgwf}) as well.  

It should be stressed that the present formalism can take into 
account all possible types of correlation in a single framework, 
so to evaluate matrix elements can be reduced to that in a 
particular representation~\cite{book}. That is, one 
does not need to calculate 
matrix elements separately in the different coordinate sets.   

The calculation of matrix elements can be performed separately 
in the orbital, spin, isospin and color parts. The permutation $P$ 
of the antisymmetrizer ${\cal A}$ causes a linear 
transformation of the coordinate ${\bfi x}$ to ${\bfi x}'$ in 
the orbital function; ${\bfi x}'\!=\!{\cal P}{\bfi x}$.
Thus the matrix element of an operator ${\cal O}$ acting in 
the orbital space can be calculated as  
\begin{equation}
\langle \phi_{LM_L}(A,u,{\bfi x})|{\cal O}P|
\phi_{L'M_L'}(A',u',{\bfi x})\rangle
=\langle \phi_{LM_L}(A,u,{\bfi x})|{\cal O}|
\phi_{L'M_L'}(\widetilde{{\cal P}}A'{\cal P},\widetilde{{\cal
P}}u',{\bfi x})\rangle.
\end{equation}
Here the action of the permutation results in just renaming 
$A'$ as $\widetilde{{\cal P}}A'{\cal P}$ and 
$u'$ as $\widetilde{{\cal P}}u'$, so that we do not 
need to change the functional form of $\phi_{LM_L}$ at all. 
The orbital matrix element was  
evaluated using the method explained in detail in~\cite{book}. 
The matrix elements involving the spin, isospin and color 
functions are evaluated using the Wigner-Eckart theorem as well as 
SU(2) and SU(3) recoupling techniques~\cite{hecht}. 

\section{Resonance in a single-channel calculation}
\label{resonance.rsm}

The mass of the $\Theta^+$ is in the continuum above the 
$N\!+\!K$ threshold. Here we briefly discuss how to predict a 
resonance mass using the basis functions 
explained in the previous section. There are some powerful 
methods such as a complex scaling method~\cite{csm} and an 
analytic continuation in a coupling constant~\cite{accc,tanaka} 
to use a generalization of bound-state problems for 
locating the resonance. We use the real stabilization 
method~\cite{hazi} among others for its simplicity. 
In the stabilization method the Schr\"odinger equation is solved 
in a box, i.e., on a square-integrable basis, which makes all 
solutions look like bound states. From among the 
bound-state-looking discrete states, those corresponding to the 
resonances are singled out by exploiting their stability 
against changes of the box size (which is practically equivalent 
to the basis dimension in the present case). 

\begin{figure}[b]
\begin{center}
\rotatebox{270}{\resizebox{6cm}{!}{\includegraphics{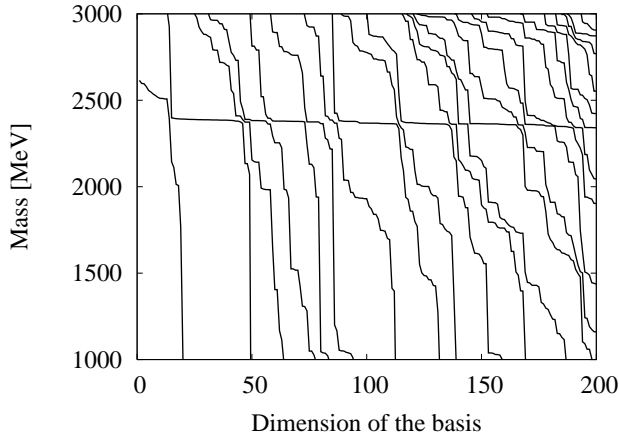}}}
\end{center}
\caption{The mass of the $\Theta^+$ with $\frac{1}{2}^+$ on a random
 basis in the Jaffe-Wilczek model. The Al1 potential is used.}
\label{random.basis}
\end{figure}

We illustrate this procedure in a single-channel calculation where 
the basis functions are limited to one particular 
spin-isospin-color configuration.  
As the single channel we adopt two representative ones: One is  
the Jaffe-Wilczek model~\cite{jaffe} in which both diquarks are 
restricted to $S\!=\!0,\, T\!=\!0,\, \Gamma\!=\!(01)$ and these
identical bosons 
are coupled to be antisymmetric in the color state, and the other  
is the Karliner-Lipkin model~\cite{karliner} in which the triquark 
with $S\!=\!\frac{1}{2},\, T\!=\!0,\, \Gamma\!=\!(10)$ 
and the diquark with 
$S\!=\!0,\, T\!=\!0,\, \Gamma\!=\!(01)$ are coupled to 
a color singlet state (see Eq.~(\ref{klmodel})). 
In the single-channel calculation, 
only the matrix $A$ and the vector $u$ are variational 
parameters characterizing the basis function, and the 
parameter space for them are first defined appropriately. 
Figure~\ref{random.basis} plots the mass eigenvalues 
vs the basis dimension for 
$J^P\!=\!\frac{1}{2}^+$ ($L\!=\!1,\, S\!=\!\frac{1}{2}$) 
in the Jaffe-Wilczek model, where 
the basis was randomly chosen from the parameter space 
without any selection procedure and just increased one by one. 
A remarkable feature of this figure is that one has a solution 
whose mass stays nearly constant. The stable solution is not a ground 
state but has a large mass above the $N\!+\!K$ threshold. The rms 
radius calculated from the stable solution is also stable: 
The mass and the rms radius are 2370 MeV and 0.72 fm for the 
basis dimension ${\cal K}$=100, and 2340 MeV and 0.73 fm for 
${\cal K}$=200, respectively. This stability 
is an indication required for the existence of a resonance.
To extend this type of calculations to a full coupled-channels 
problem is hard, however, because the number of channels is 
fairly large.  
It is advisable to confirm the stability even though the basis 
dimension is truncated through some selection procedure. 

\begin{figure}[t]
\begin{center}
\rotatebox{270}{\resizebox{4.8cm}{!}{\includegraphics{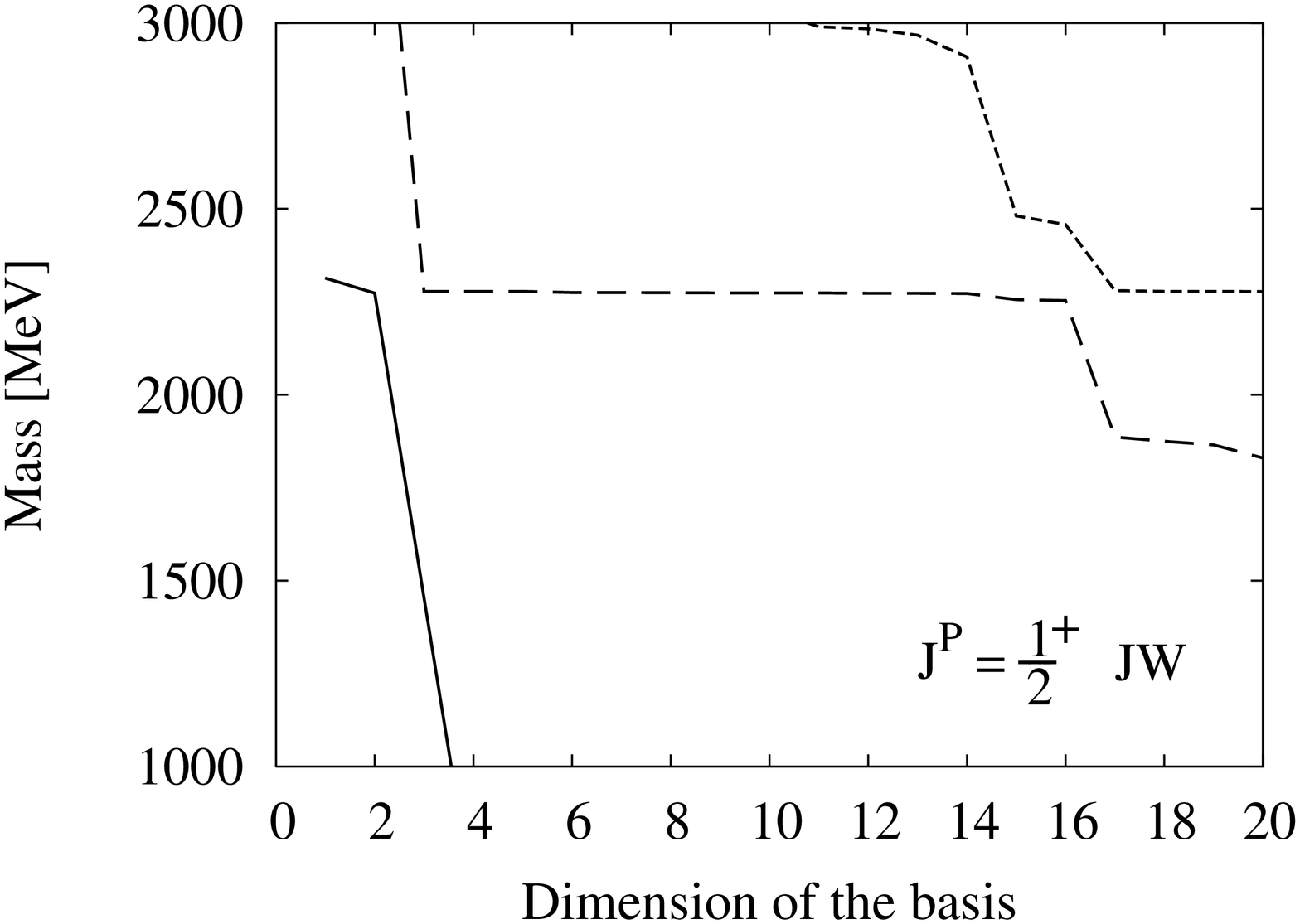}}}
\hspace{-0.8cm}
\rotatebox{270}{\resizebox{4.8cm}{!}{\includegraphics{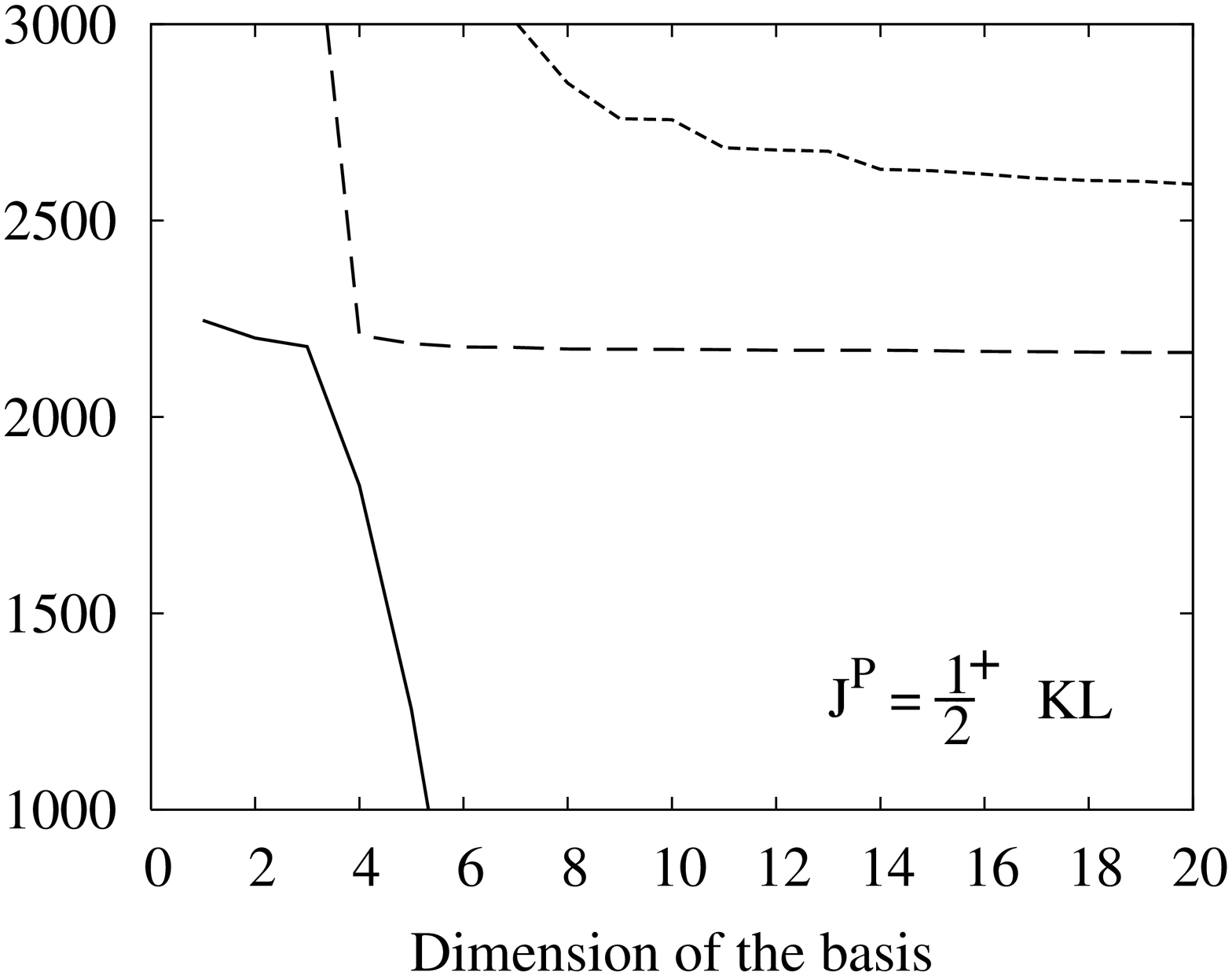}}}
\\
\hspace*{0.28cm}
\rotatebox{270}{\resizebox{4.8cm}{6.5cm}{\includegraphics{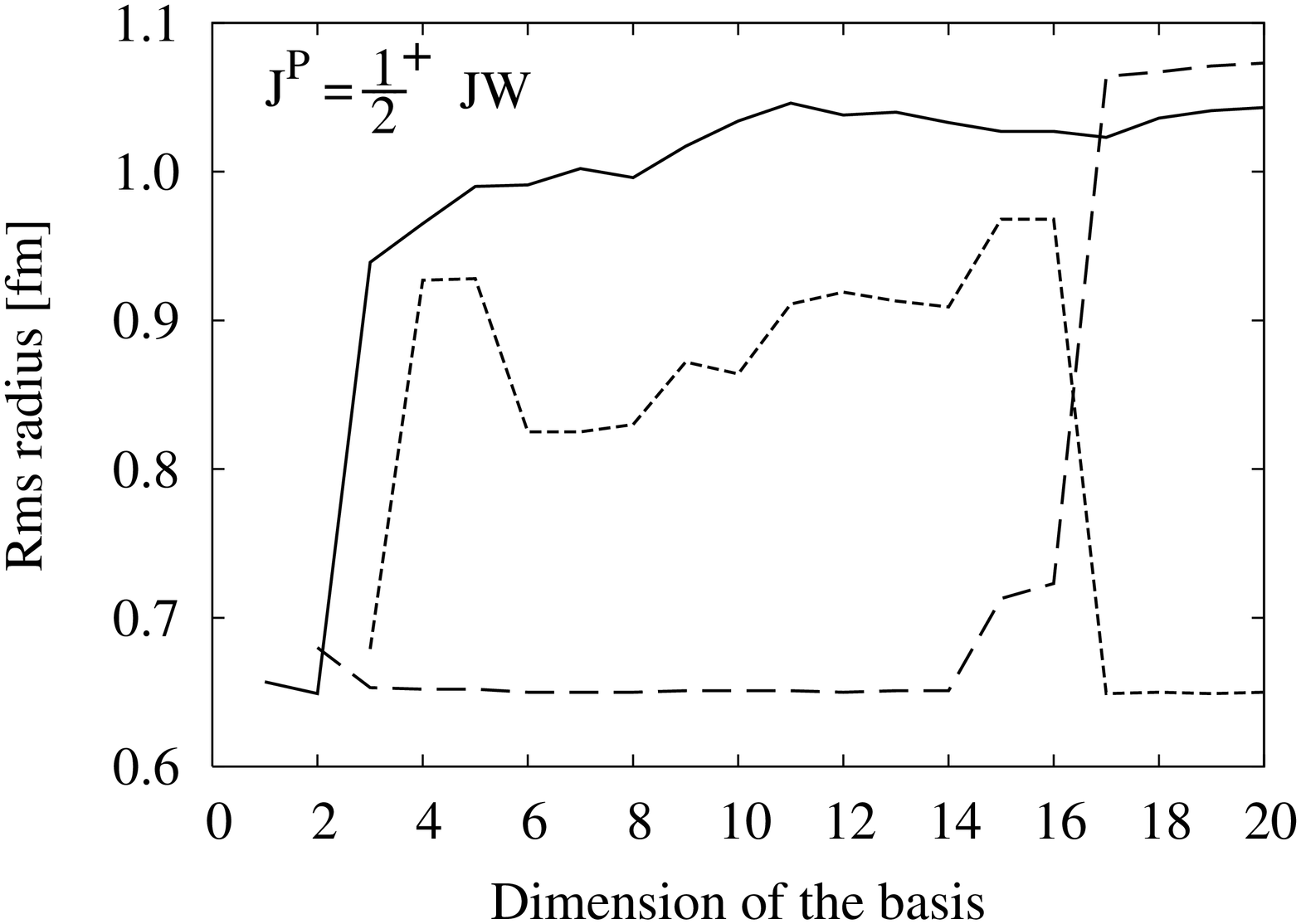}}}
\hspace{-0.42cm}
\rotatebox{270}{\resizebox{4.8cm}{6.5cm}{\includegraphics{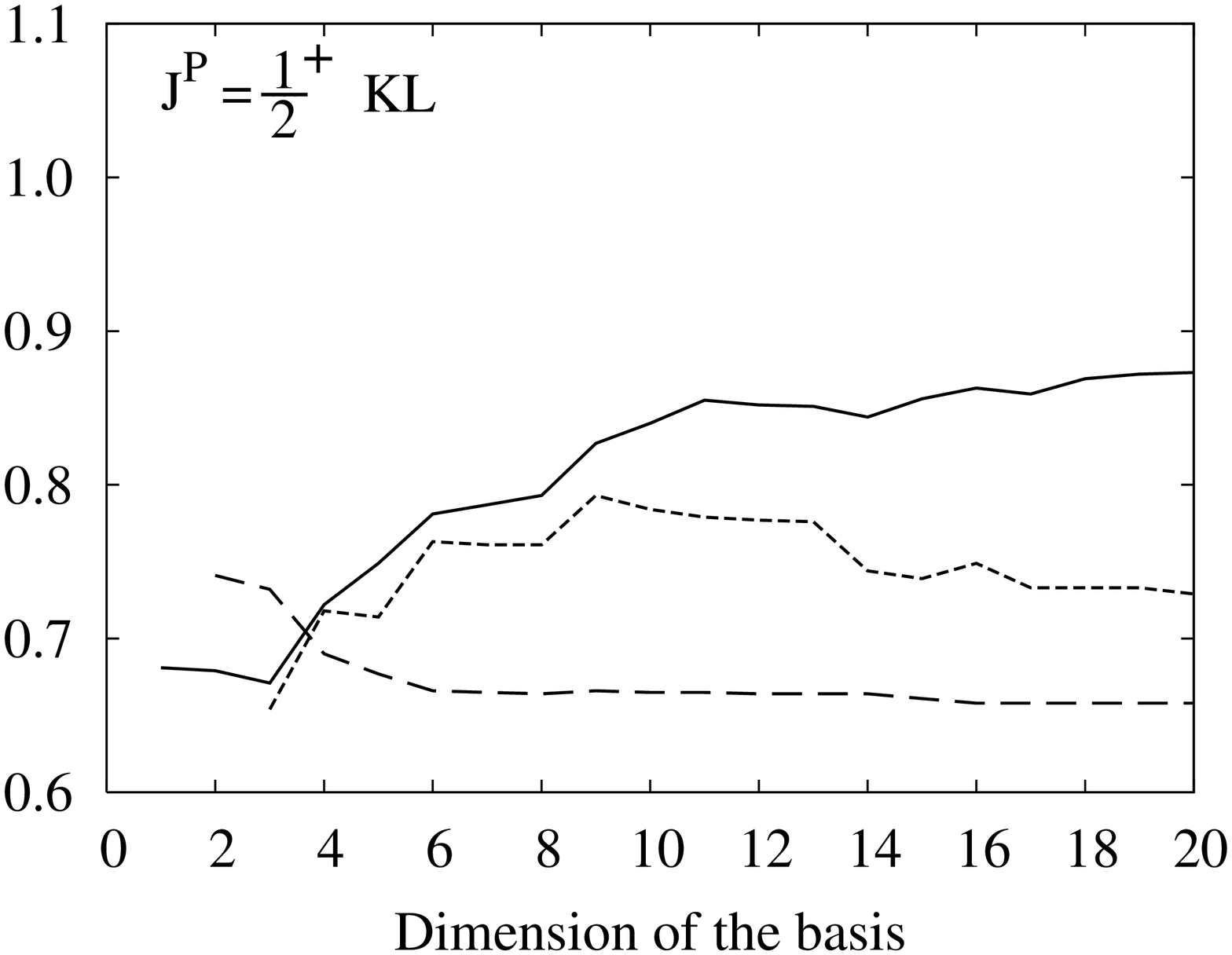}}}
\end{center}
\caption{The mass and rms radius of the $\Theta^+$ 
in a single-channel calculation. 
The figures on the left side are the results 
in the Jaffe-Wilczek model, while those on the right side
are obtained in the Karliner-Lipkin model. The AL1 potential 
is used.}
\label{fig.2}
\end{figure}

Examples of such a truncated calculation are presented 
in Fig.~\ref{fig.2} where the mass as well 
as the rms radius are displayed as a function of the basis 
dimension. Here 
the basis selection was performed using 
the SVM~\cite{svm,book} as it leads to virtually exact 
solutions in rather small dimension and its  
power is proved in diverse few-body problems.  
In the SVM the basis dimension is increased one by one 
by testing a number of candidates which are chosen randomly.  
The importance of 
each candidate is evaluated by calculating the energy gain 
produced when it is included in the basis set. The candidate 
which produces the lowest of the energies is chosen to be 
a member of the basis set. Furthermore, it is shown 
in~\cite{usukura} 
that the basis set for describing a resonance can be selected 
very efficiently with the SVM. The basis selection in 
the present calculation was actually done as follows: each 
element of 
$A$ and $u$ was randomly generated twenty times, and the best one 
among these trials was selected as a tentative candidate 
for a new basis. This process was repeated ten times to select 
a most successful one from among these tentative candidates. A 
typical computer time needed to determine twenty basis 
states was about half an hour on a computer loading the 
Intel (R) Pentium (R) 4 CPU 3.20 GHz. 

We see two characteristics in either model of the figure. 
One is that the lowest mass decreases as the basis dimension 
increases, reaching even a negative value. The other is the 
appearance of such a unique mass eigenvalue that is stable 
against the increase of the basis dimension. By looking at the 
figure of the rms radius, we note that the radius corresponding to 
the former solution is rather large and continues to increase 
with increasing dimension. In contrast to this case, the radius 
for the latter solution stays approximately constant, and it 
decreases a little from the one of Fig.~\ref{random.basis}. 
According to the criterion of the real stabilization 
method, we may identify the latter mass as the resonance mass 
and the corresponding wave function as the approximate 
resonance wave function which is valid except at large distances. 
Of course the mass obtained here is just 
a result of the single-channel calculation, so before declaring 
it to be the $\Theta^+$ mass we have to consider the effect of 
other channels which may contribute to the resonance. 
In particular, the effect of the $NK$ channel must be examined 
carefully. This 
will be performed in the next section. 

We have performed a single-channel calculation for the other 
configurations as well in each case of the successive coupling, 
the diquark model or the baryon-meson model. The falloff 
of the mass and the increase 
of the rms radius are observed in all the cases. The reason for 
this seems due to the color dependence of the quark-quark 
potential as was pointed out in sect.~\ref{hamiltonian}. 
First we notice that the mass of the colored subsystem 
happens to be negative and its radius gets very large obviously 
because of the 
$(\lambda_i^{\rm C}\cdot\lambda_j^{\rm C})r_{ij}$ operator 
which works to deconfine the system in special color channels. 
A five-quark system usually contains such colored components, so 
the variational ground state will make use of it to gain the 
energy. 

It is nevertheless fortunate that we have the sign of a 
resonance, that is, the stability in the mass diagram, 
which is indispensable for getting a resonance in the 
bound-state-looking calculations. Those cases which show the 
stability 
are characterized by that at least one of the diquarks is 
antisymmetric with respect to the simultaneous interchange of 
the spin, isospin and color degrees of freedom. 

A single-channel calculation has also been performed for 
other $J^P$ states. We have obtained some stable masses around 
2500, 2700 and 2900 MeV for 
$(L\!=\!1,\, S\!=\!\frac{3}{2})$, $(L\!=\!2,\, S\!=\!\frac{1}{2})$ and 
$(L\!=\!2,\, S\!=\!\frac{3}{2})$ 
cases, respectively. It seems that the larger $L$ and the larger 
$S$ the system has the larger its mass is. This can be expected 
from the role played by the kinetic energy and the color 
magnetic piece of the quark-quark potential. We will focus 
our attention 
on the $J^P\!=\!\frac{1}{2}^{\pm}$ and $\frac{3}{2}^-$ states 
in what follows.

\section{Results}
\label{results}

\subsection{Mass spectrum}
\label{mass.spectrum}

We have made a coupled-channels calculation to predict 
the mass of the $\Theta^+$ for $J^P\!=\!\frac{1}{2}^-\, 
(L\!=\!0,S\!=\!\frac{1}{2},T\!=\!0)$, 
$\frac{1}{2}^+(\frac{3}{2}^+)\, (L\!=\!1,S\!=\!\frac{1}{2},T\!=\!0)$ 
and $\frac{3}{2}^-\, (L\!=\!0,S\!=\!\frac{3}{2},T\!=\!0)$. 
As discussed in sect.~\ref{wavefunction}, 
there are in principle 30 channels 
for $S\!=\!\frac{1}{2},\, T\!=\!0$ and 24 channels for 
$S\!=\!\frac{3}{2},\, T\!=\!0$. Among these some channels 
play an important role to produce a resonance 
but some others play much less significant roles. In the actual
calculation we first single out basis functions from those 
channels which give a stable mass in a single-channel 
calculation, and include them in the 
coupled-channels calculation. Included are 7, 5 and 12 channels 
for $J^P\!=\!\frac{1}{2}^-, \frac{3}{2}^-$ 
and $\frac{1}{2}^+(\frac{3}{2}^+)$, respectively. These 
channels are expressed using the channel labels of  
$S_{12}, T_{12}, \Gamma_{12}, S_{34}, T_{34}, \Gamma_{34}, 
S_{1234}$ in the diquark model as follows:
\begin{eqnarray}
{\rm for\ }J^P&=&\textstyle{\frac{1}{2}}^- \ \ \ \ \ 00(01)00(01)0, \ 
00(01)10(20)1, \ 01(20)11(01)1, \ 11(01)11(01)0, 
\nonumber \\
& &\ \, \hspace*{1cm} 10(20)00(01)1, \ 11(01)01(20)1, 
\ 11(01)11(01)1, 
\nonumber \\
{\rm for\ }J^P&=&\textstyle{\frac{3}{2}}^- \ \ \ \ \  10(20)00(01)1, \ 
11(01)01(20)1, \ 00(01)10(20)1, \ 01(20)11(01)1, 
\nonumber \\
& &\ \, \hspace*{1cm} 11(01)11(01)1,
\nonumber \\
{\rm for\ }J^P&=&\textstyle{\frac{1}{2}}^+ \ \ \ \ \ 00(01)00(01)0, \ 
00(01)10(20)1, \ 01(20)11(01)1, \ 11(01)11(01)0, 
\nonumber \\
& &\ \, \hspace*{1cm} 10(20)00(01)1, \ 
11(01)01(20)1, \ 10(01)10(20)1, \ 10(20)10(01)1, 
\nonumber \\
& &\ \, \hspace*{1cm} 11(01)11(01)1, \ 11(20)11(01)1, 
\ 10(01)00(01)1, \ 00(01)10(01)1.
\end{eqnarray}

To obtain a final 
result for a resonance, we further consider the effect of those 
basis functions which are generated from the $NK$ channel 
for $J^P\!=\!\frac{1}{2}^{\pm}$ or from the $NK^*$ channel for 
$\frac{3}{2}^-$ because the coupling of the resonance with 
that channel appears to be important for calculating the width. 
Note that the $NK$ channel does not couple to $\frac{3}{2}^-$ 
states as they differ in the total spin $S$, but the 
$NK^*$ channel can have the same spin $S\!=\!\frac{3}{2}$. 
For example, the $NK$ channel basis functions included 
are expressed as 

\begin{equation}
\Phi_i(NK)= {\cal A}\Bigg\{[[{\Psi}_{\frac{1}{2}\frac{1}{2}}(N)
\Psi_{0\frac{1}{2}}(K)]_{I=\frac{1}{2}\, T=0M_T=0}
\exp\Big(-\frac{1}{2}\,a_i{\bfi z}_4^{\, 2}\Big){\cal Y}_{\ell}({\bfi z}_4)]_{JM}\Bigg\},
\label{nkbasiswf}
\end{equation}
where ${\Psi}_{I=\frac{1}{2}\, T=\frac{1}{2}}(N)$ and 
$\Psi_{I=0\, T=\frac{1}{2}}(K)$ are respectively 
the wave functions of $N$ and $K$ which are obtained by 
solving the three-quark and quark-antiquark Hamiltonians with 
the potential of Eq.~(\ref{al1pot}) or Eq.~(\ref{tspot}). 
These are coupled to the relative motion function of 
Gaussian form with the 
orbital angular momentum $\ell$ to the total angular momentum 
$JM$. Here the coordinate 
${\bfi z}_4\!=\!{\bfi R}_N\!-\!{\bfi R}_K$ is the 
relative distance vector between the center of mass of $N$, 
${\bfi R}_N$, and the center of mass of $K$, ${\bfi R}_K$. 
In the calculation the length parameter 
is set to be $1/{\sqrt{a_i}}=0.6\times 1.4^{i-1}\, {\rm fm}\ 
(i=1,2,\ldots,5)$.

\begin{figure}[b]
\begin{center}
\begin{tabular}{cc}
\rotatebox{270}{\resizebox{6.5cm}{7cm}{\includegraphics{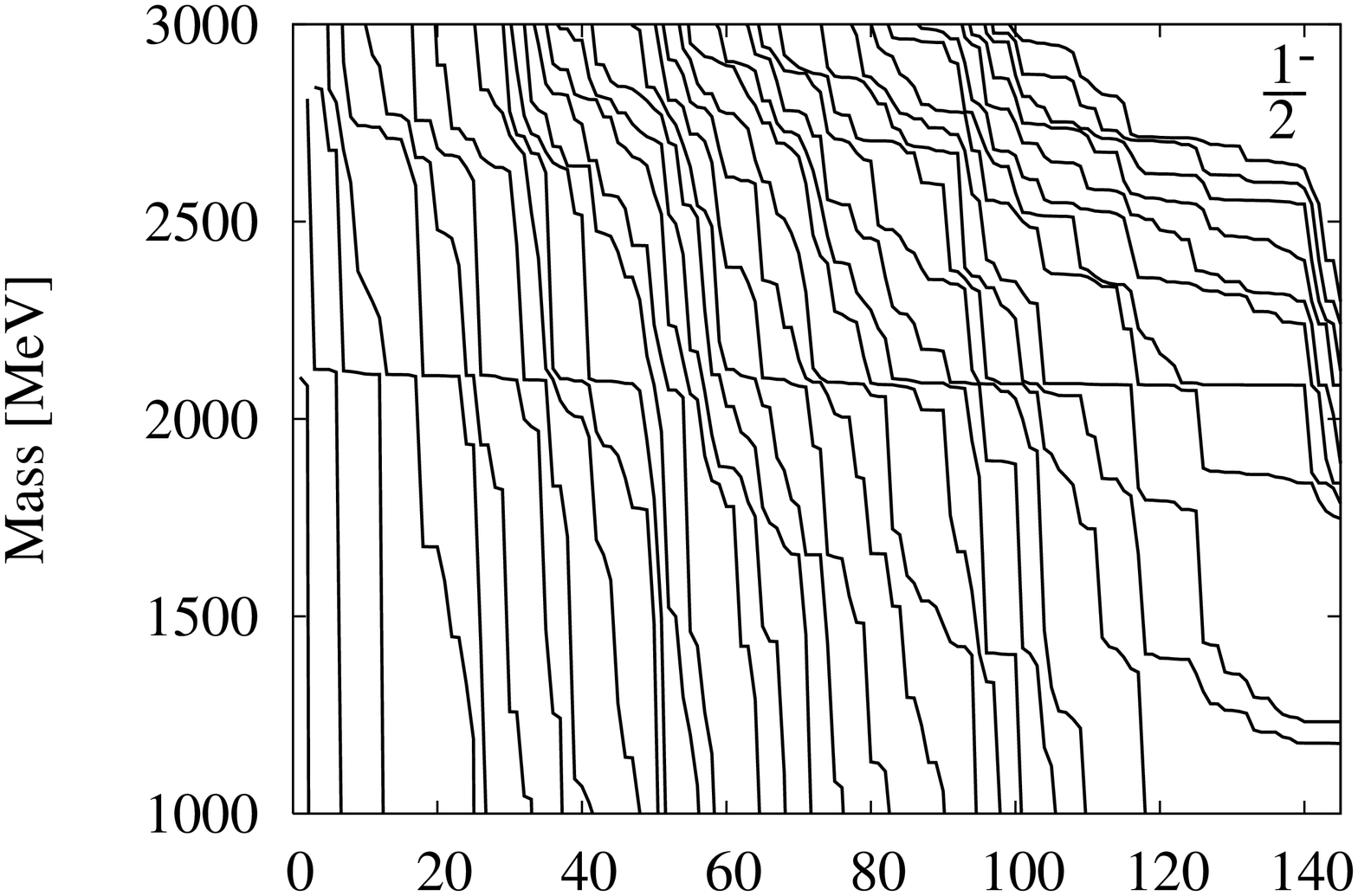}}}
\hspace{-0.8cm}
\vspace{-1cm}
\rotatebox{270}{\resizebox{6.5cm}{7cm}{\includegraphics{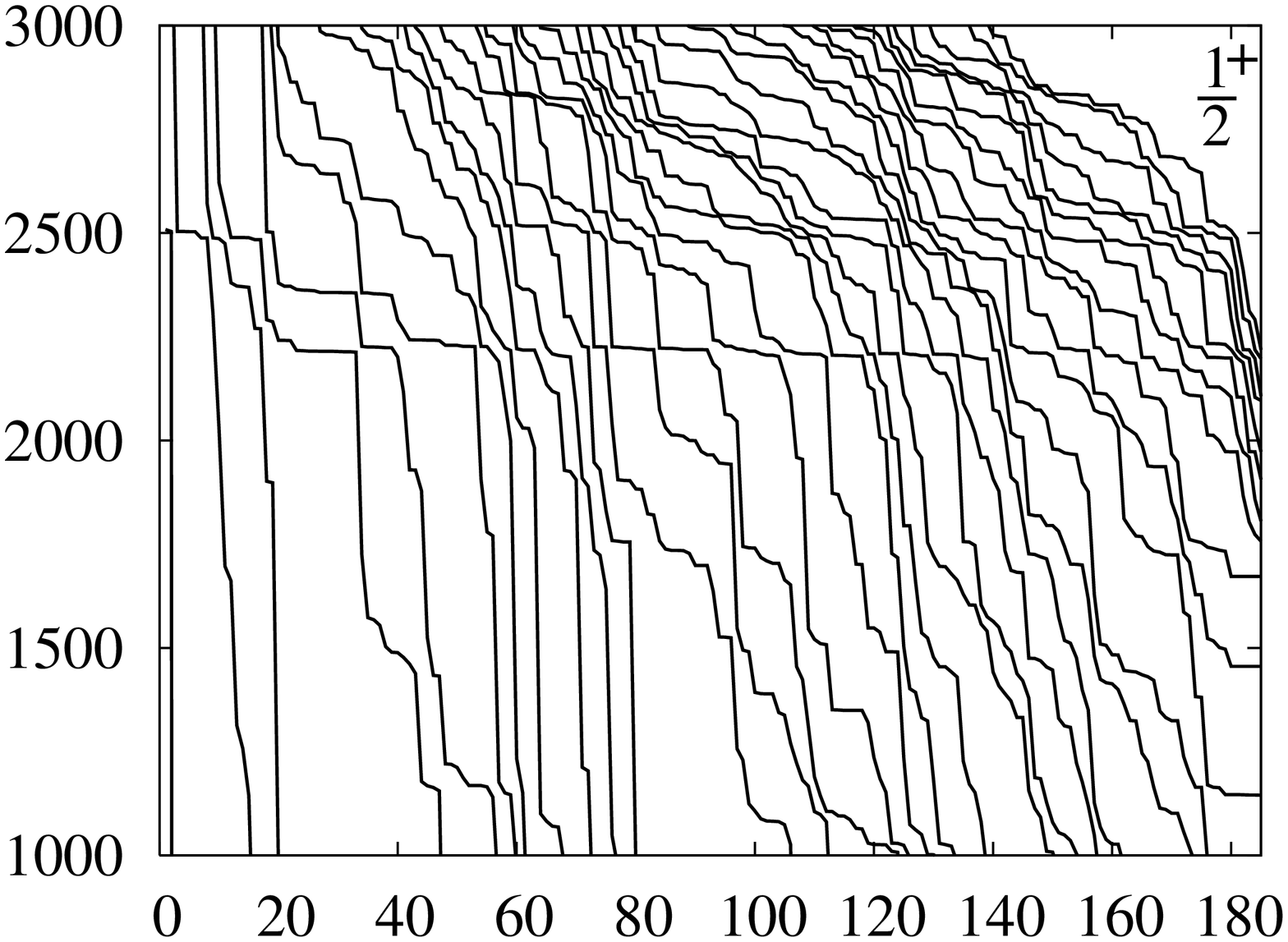}}}\\
\hspace{-6.5cm}
\rotatebox{270}{\resizebox{6.5cm}{7cm}{\includegraphics{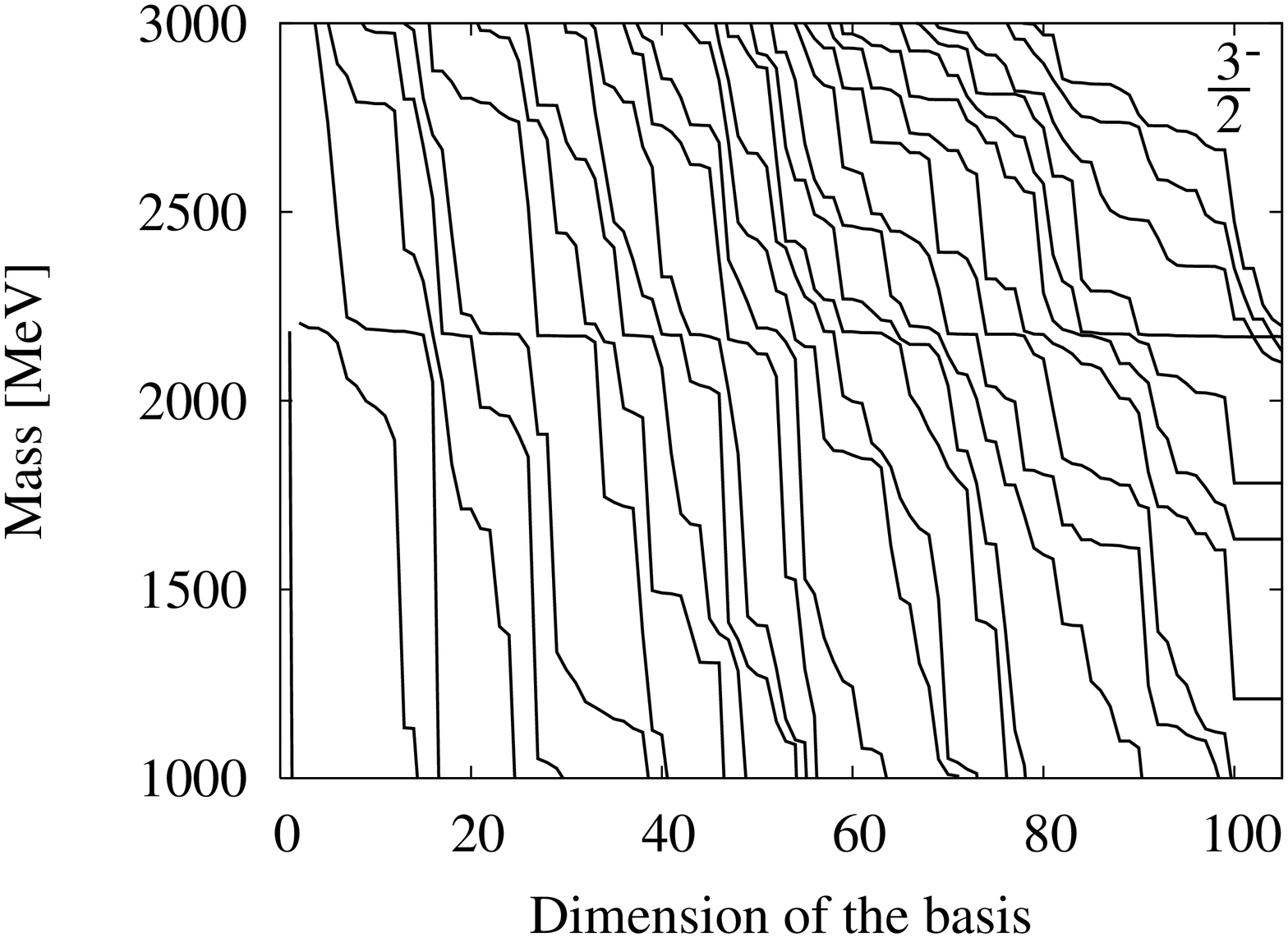}}}\\
\end{tabular}
\end{center}
\caption{The mass of the $\Theta^+$ in a coupled-channels calculation. 
The $NK$ channel is explicitly included for 
$J^P\!=\!\frac{1}{2}^{\pm}$, while the $NK^*$ channel is 
for $\frac{3}{2}^-$. The AL1 potential is used. }
\label{al1spectra}
\end{figure}

\begin{figure}[thb]
\begin{center}
\begin{tabular}{cc}
\rotatebox{270}{\resizebox{6.5cm}{7cm}{\includegraphics{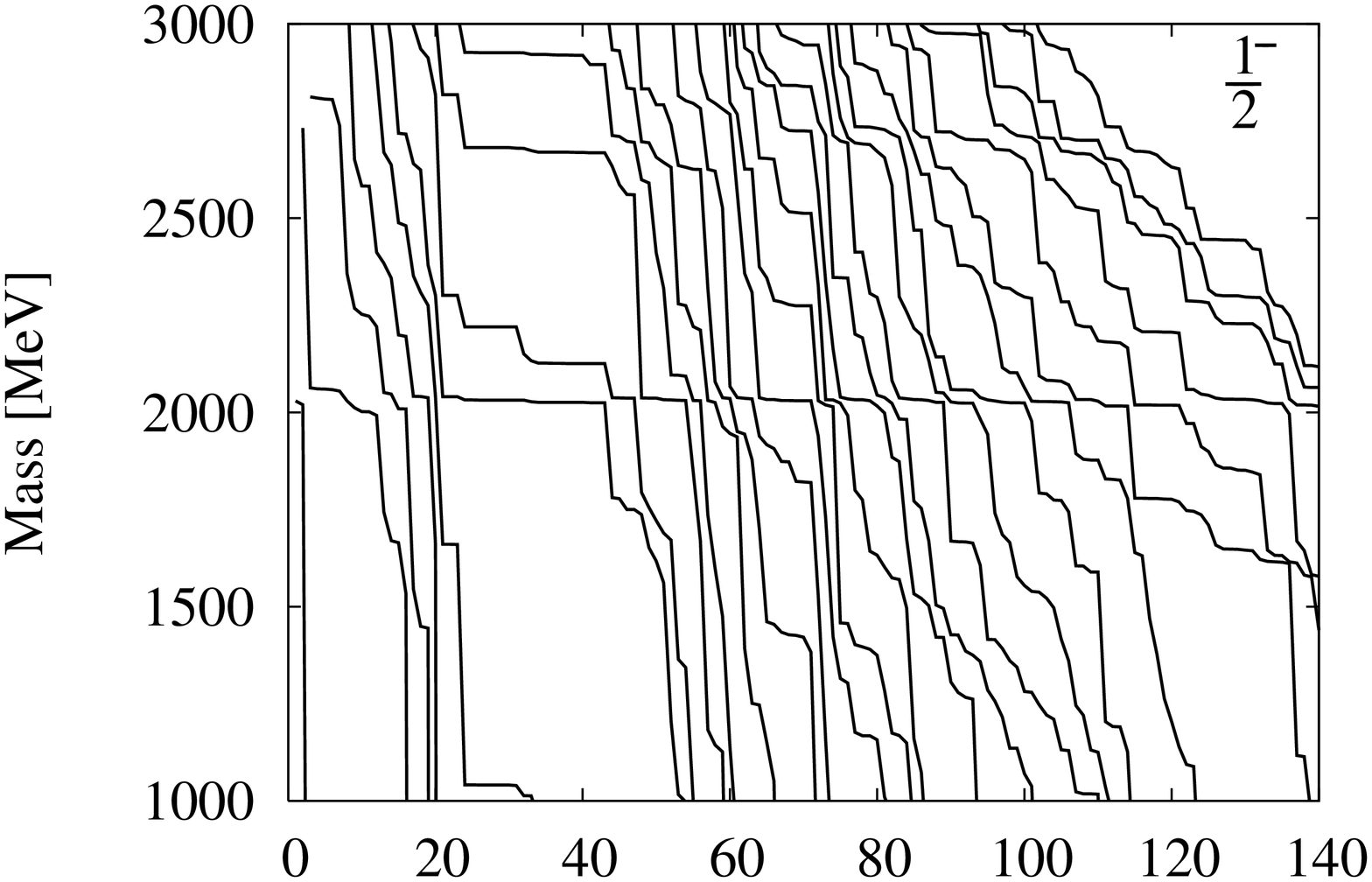}}}
\hspace{-0.8cm}
\vspace{-1cm}
\rotatebox{270}{\resizebox{6.5cm}{7cm}{\includegraphics{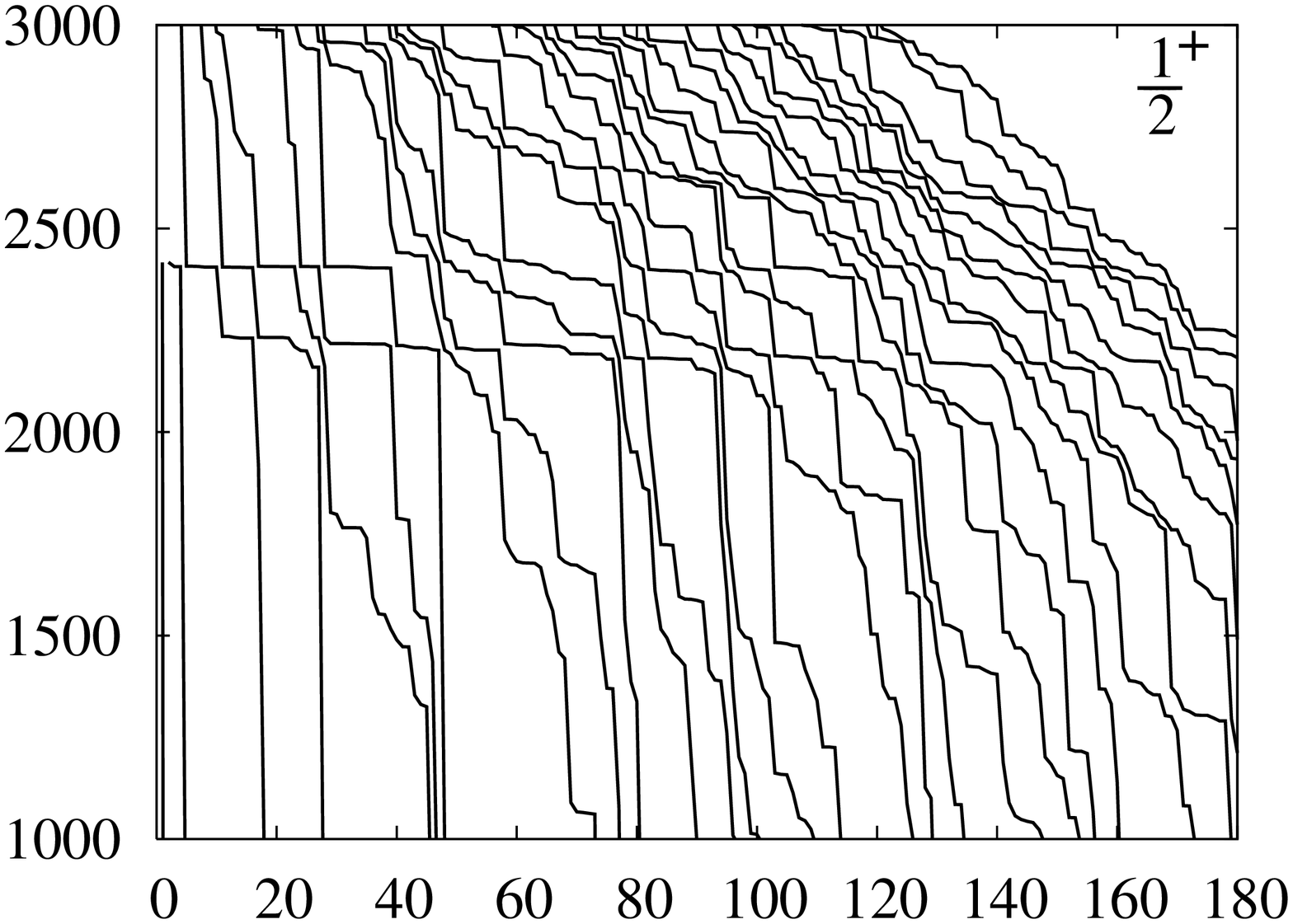}}}\\
\hspace{-6.5cm}
\rotatebox{270}{\resizebox{6.5cm}{7cm}{\includegraphics{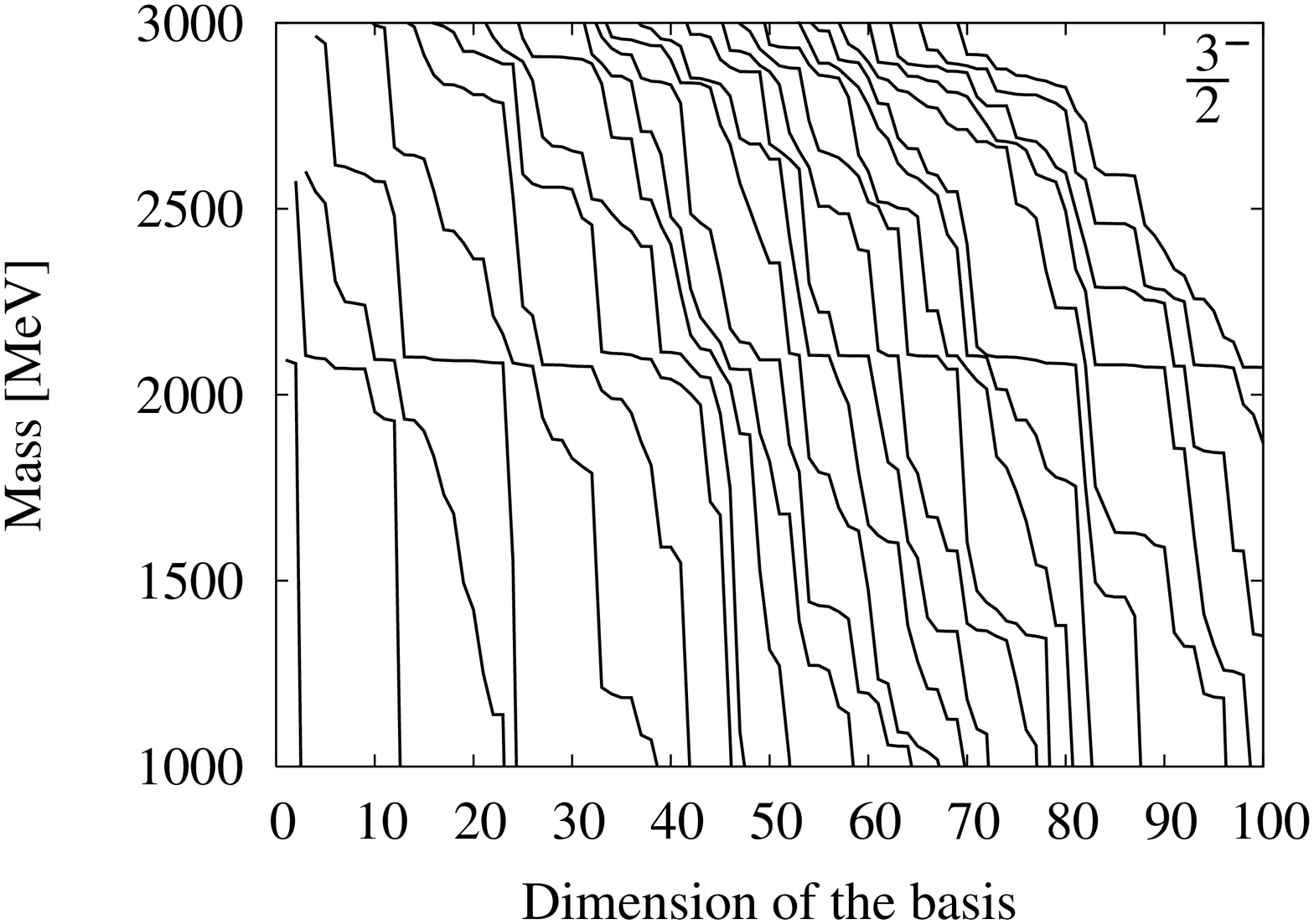}}}\\
\end{tabular}
\end{center}
\caption{The same as Fig.~\ref{al1spectra} but for the TS potential. }
\label{tsspectra}
\end{figure}

The Hamiltonian is diagonalized in the set of the 
chosen basis functions. Figures~\ref{al1spectra} and ~\ref{tsspectra} 
display the calculated mass  
as a function of the basis dimension. We 
clearly see a stabilized mass in the diagram and have confirmed that 
the rms radius calculated from the 
corresponding wave function is stable. This stabilized solution 
is accepted as a resonance we seek. We have found that the mass and 
the radius of the stabilized solution remains basically unchanged 
by including the $NK$ (and $NK^*$) channel of 
Eq.~(\ref{nkbasiswf}). Evidently, our resonance 
does not correspond to a variational energy minimum. We have used 
a variational method in order to set up compact, 
appropriate basis functions. This is in 
contrast to the approach in~\cite{takeuchi,enyo}, 
where a variational energy minimum is looked for and moreover 
a coupling to the baryon-meson channel like $NK$ is 
excluded from the model space. 

\begin{figure}[t]
\begin{center}
\begin{tabular}{cc}
\rotatebox{270}{\resizebox{6.5cm}{!}{\includegraphics{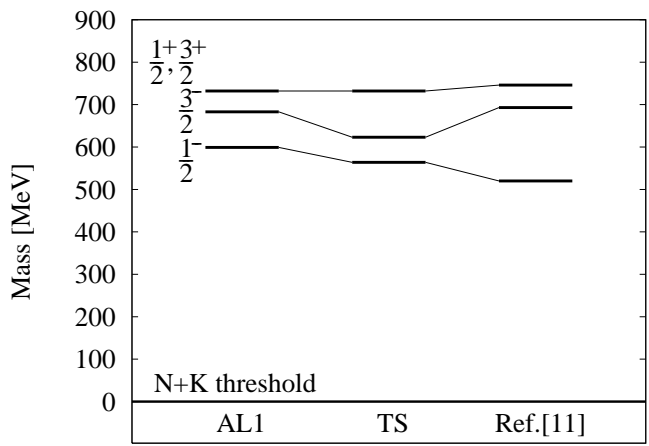}}}
\end{tabular}
\end{center}
\caption{The mass spectrum of the $\Theta^+$ from the 
$N\!+\!K$ threshold. The result of Ref.~\cite{takeuchi} is due to 
the TS potential. }
\label{mass.spectra}
\end{figure}

The masses and rms radii obtained in the 
present calculation are summarized in Fig.~\ref{mass.spectra} 
and Table~\ref{mass.radius-table}. 
It is concluded that the mass of the $\Theta^+$ increases 
in order of $J^P=\frac{1}{2}^-, \frac{3}{2}^-,
\frac{1}{2}^+(\frac{3}{2}^+)$, which agrees with the result of
Ref.~\cite{takeuchi}. A lower mass is obtained for the $\frac{1}{2}^-$ state 
than for the $\frac{1}{2}^+$ state, which is consistent with the recent 
variational Monte carlo calculation~\cite{paris} but is opposite to 
the level order obtained in \cite{hiyama}. 
The color magnetic and kinetic energy terms in the Hamiltonian play 
the most important contribution to generate the mass differences. 
The mass splitting between the $\frac{3}{2}^-$ and 
$\frac{1}{2}^+(\frac{3}{2}^+)$ states is smaller in the AL1 
potential than that in the TS potential, but both 
potentials otherwise give qualitatively similar results. 
Our result for the TS potential is 
compared to that of \cite{takeuchi}, and we see that 
the mass splitting is not necessarily the same between the two.  
This is probably due to the difference in 
obtaining the mass as well as in specifying the correlation of the 
quark dynamics.

\begin{table}[b]
\caption{The mass and rms radius of the $\Theta^+$. The calculated 
$N\!+\!K$ threshold energy is 1486 MeV for the AL1 potential and 1451 MeV for 
the TS potential, respectively.  }
\begin{center}
\begin{tabular}{cccccc}
\hline\hline
 & \multicolumn{2}{c}{AL1} & & \multicolumn{2}{c}{TS}\\
\cline{2-3}\cline{5-6}
$J^P$  & $M$ [MeV] & $\sqrt{\langle r^2\rangle}$ [fm]
& & $M$ [MeV] & $\sqrt{\langle r^2\rangle}$ [fm]\\
\hline
$\frac{1}{2}^-$ & 2086    & 0.57 & & 2015 & 0.63\\
$\frac{3}{2}^-$ & 2169    & 0.59 & & 2074 & 0.63\\
$\frac{1}{2}^+,\frac{3}{2}^+$ & 2219    & 0.74 & & 2183 & 0.72\\
\hline\hline
\end{tabular}
\end{center}
\label{mass.radius-table}
\end{table}

The calculated mass of the $\Theta^+$ is higher than the observed
mass ($\sim$1540 MeV) by about 500-600 MeV. This appears to be 
common in all the calculations using the potential of OGE 
type~\cite{takeuchi,enyo,hiyama} or the OGE plus one-pion 
exchange potential~\cite{paris}. 
The zero-point energy 
term is uncertain, so it is hard to get an absolute 
mass unambiguously. There are some discussions to reduce the 
mass. One is to consider the flavor-spin dependence of the 
quark-quark potential or the one-boson exchange 
potential~\cite{stancu,carlson}. 
The second is to take into account the role of the 
instanton induced interaction~\cite{shuryak,kochelev}. Though 
this interaction is expected to bring about an extra attraction, 
its effect has not yet been quantified. The third is to consider 
the relativistic effect of the quark motion. An extensive 
calculation has been made in~\cite{takeuchi} which used 
the semirelativistic expression for the kinetic energy and the 
one-boson exchange potential as well as the OGE 
potential. It seems that the $\frac{1}{2}^+$ state comes down 
close to or even lower than the $\frac{3}{2}^-$ state. The 
calculated mass is, however, still larger than the observed 
mass, and even goes around 
2 GeV if the zero-point energy is taken to be proportional to 
the number of quarks.

\subsection{Decay width and spin-parity}
\label{decay}

The width of a resonance is in principle calculated through a 
coupling to the continuum states of decay channels. 
As we have localized the $\Theta^+$, 
it may be appealing to use, for example, the complex scaling 
method to calculate the width as was done in~\cite{usukura}. 
This is, however, too sophisticated in the present case since the 
calculated $\Theta^+$ mass is subject to large uncertainty 
of the quark-quark potential. On the other hand, the wave function of the 
resonance remains the same for an arbitrary adjustment of the 
resonance energy due to the change of the zero-point 
energy term, so we instead use the $R$-matrix 
theory~\cite{r-matrix,horiuchi} in which the width $\Gamma$ of 
an isolated resonance is calculated through
\begin{equation}
\Gamma=2P_{\ell}(a)\gamma^2(a), 
\label{decay.Rmatrix}
\end{equation}
where $a$ is a channel radius and $P_{\ell}(a)$ is the 
penetrability given by
\begin{equation}
P_{\ell}(a)=\frac{ka}{j_{\ell}^{\, 2}(ka)+n_{\ell}^{\, 2}(ka)}.
\end{equation}
Here $\ell$ and $k$ are the orbital angular momentum and the 
wave number of the relative motion between the decaying 
particles, and $j_{\ell}$ and $n_{\ell}$ are the spherical Bessel functions. 
When both of the decaying particles are charged, the spherical 
Bessel functions should be replaced by the Coulomb functions. 

Most crucial in Eq.~(\ref{decay.Rmatrix}) is the reduced width 
$\gamma^2(a)$, which is related to the reduced width 
amplitude $y(a)$ for the decay to a baryon $B$ and a meson $M$:
\begin{eqnarray}
\gamma^2(a)&=&\frac{a}{2\mu }y^2(a),
\nonumber \\
y(r) &=&
\sqrt{\frac{4!}{3!}}\Bigg\langle\Big[[\Psi_{I_BT_B}(B)
\Psi_{I_MT_M}(M)]_{I\, TM_T} Y_{\ell}(\widehat{\bfi z_4})\Big]_{JM}
\frac{\delta(z_4-r)}{z_4 r}\Bigg|\Psi_{JM}^P\Bigg\rangle.
\end{eqnarray}
Here $\mu$ is the reduced mass of the baryon and the meson, 
$\Psi_{I_BT_B}(B)$ is the baryon wave function which is 
properly antisymmetrized and normalized, and 
$\Psi_{I_MT_M}(M)$ is the normalized meson wave function. 
The coordinate 
${\bfi z}_4$ is the relative distance vector between the 
baryon's center of mass and the meson's center of mass. 
Introducing a complete basis set $\{f_{n\ell}(r)Y_{\ell m}
(\hat{\bfi r})\}$, the calculation of $y(r)$ is reduced to 
that of the overlap:
\begin{equation}
y(r) =
\sqrt{\frac{4!}{3!}}\sum_{n}f_{n\ell}(r)
\Big\langle\Big[[\Psi_{I_BT_B}(B)
\Psi_{I_MT_M}(M)]_{I\, TM_T} f_{n\ell}(z_4) Y_{\ell}(\widehat{\bfi z_4})\big]_{JM}
\Big|\Psi_{JM}\Big\rangle.
\end{equation}

\begin{figure}[b]
\begin{center}
\begin{tabular}{cc}
\rotatebox{270}{\resizebox{7.5cm}{!}{\includegraphics{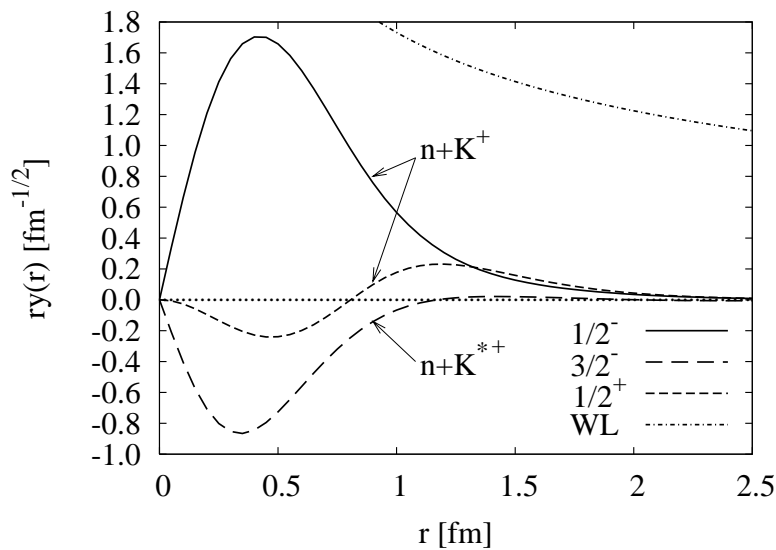}}}
\end{tabular}
\end{center}
\caption{The reduced width amplitude of the $\Theta^+$ 
to the $nK^+$ channel for $J^P=\frac{1}{2}^{\pm}$  and to the 
$nK^{*+}$ channel for $J^P=\frac{3}{2}^{-}$. The reduced 
width amplitude for $J^P=\frac{3}{2}^{+}$ is the same as that for 
$J^P=\frac{1}{2}^{+}$. The AL1 potential 
is used.}
\label{rwa.al1} 
\end{figure}

Figure~\ref{rwa.al1} displays the reduced width amplitude of the 
$\Theta^+$ resonance obtained with the AL1 potential. 
The spins of the decay channel 
are taken as $(I,\ell)\! =\! 
(\frac{1}{2},0)$ for $J^P\!=\!\frac{1}{2}^-$ and 
$(I,\ell)\! = \!(\frac{1}{2}, 1)$ for $J^P\!=\!\frac{1}{2}^+
(\frac{3}{2}^+)$, respectively. Note that the $\Theta^+$ with 
$J^P\!=\!\frac{3}{2}^-$ cannot decay to the $nK^+$ channel 
because 
it has $S\!=\!\frac{3}{2}$ and its reduced width amplitude 
to that channel vanishes. For it to decay to the $nK^+$ channel, 
one has to take into account tensor components of the 
interaction between 
the quarks. However, the $\frac{3}{2}^-$ state can decay 
to the $nK^{*+}$ channel, so the reduced width amplitude for 
this decay with $(I,\ell)\! =\! (\frac{3}{2}, 0)$ 
is displayed in the figure. 
It is found that the reduced width amplitude hardly changes 
with the inclusion of the $NK$ (or $NK^*$) 
channel basis functions defined in Eq.~(\ref{nkbasiswf}). The amplitude 
of the Wigner limit (WL), $y(r)\!=\!\sqrt{\frac{3}{r^3}}$, is also 
plotted in the figure. The Wigner limit is the partial width of a 
particular state whose wave function is uniform up to $r$ and 
zero beyond, and it is a measure to judge whether the resonance 
has a large component in the decay channel or not. 
Since the calculated reduced width amplitudes are considerably 
small compared to the Wigner limit, we can conclude that the 
$\Theta^+$ does not have large $nK^+$ component. 
This point will be confirmed in sect.~\ref{structure}. 
It is noted that the amplitude of the $\frac{3}{2}^-$ state 
is especially small at $r \geq 1$ fm. 

\begin{figure}[b]
\begin{center}
\begin{tabular}{cc}
\rotatebox{270}{\resizebox{5.5cm}{!}{\includegraphics{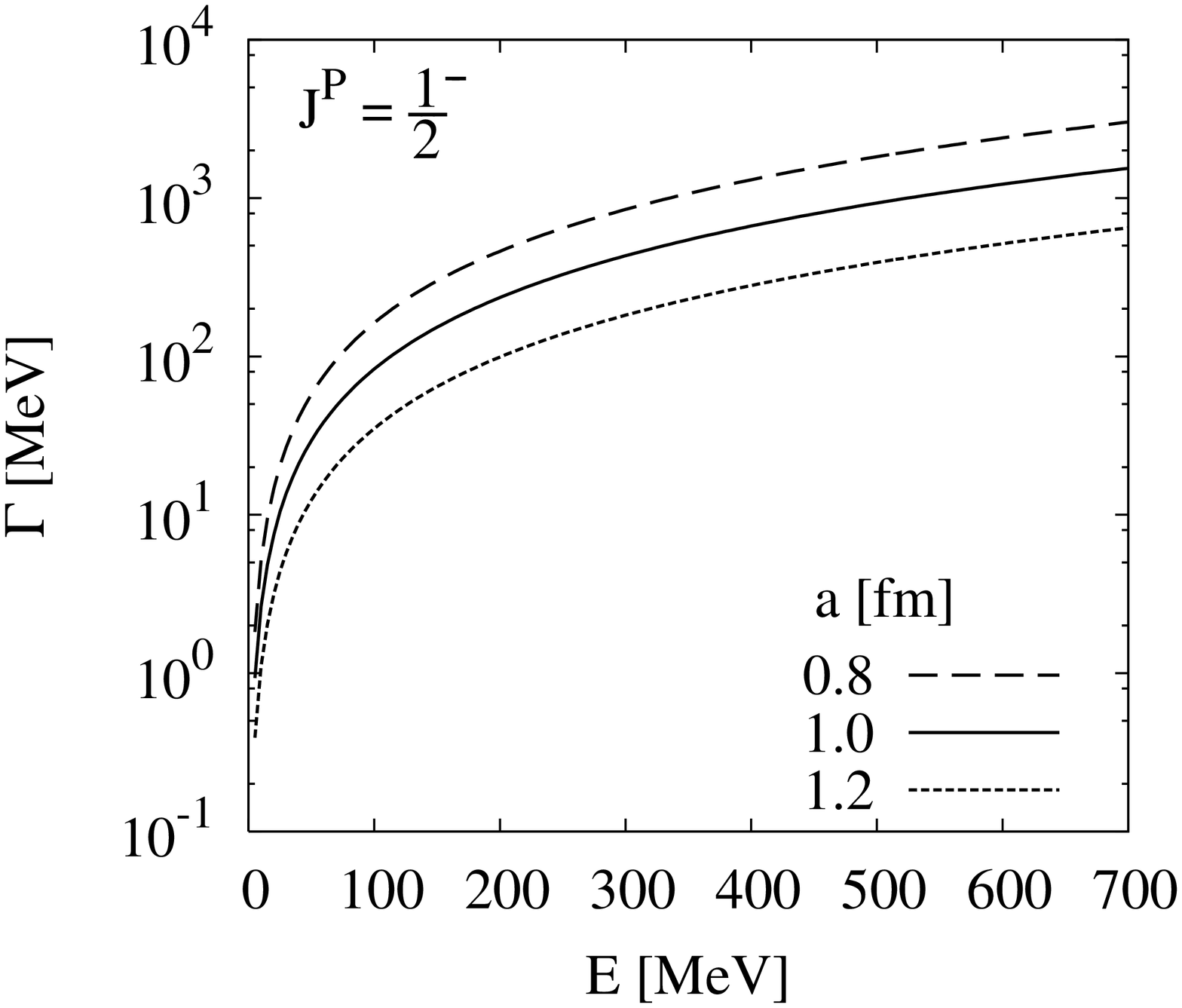}}}
\hspace{-1.4cm}
\rotatebox{270}{\resizebox{5.5cm}{!}{\includegraphics{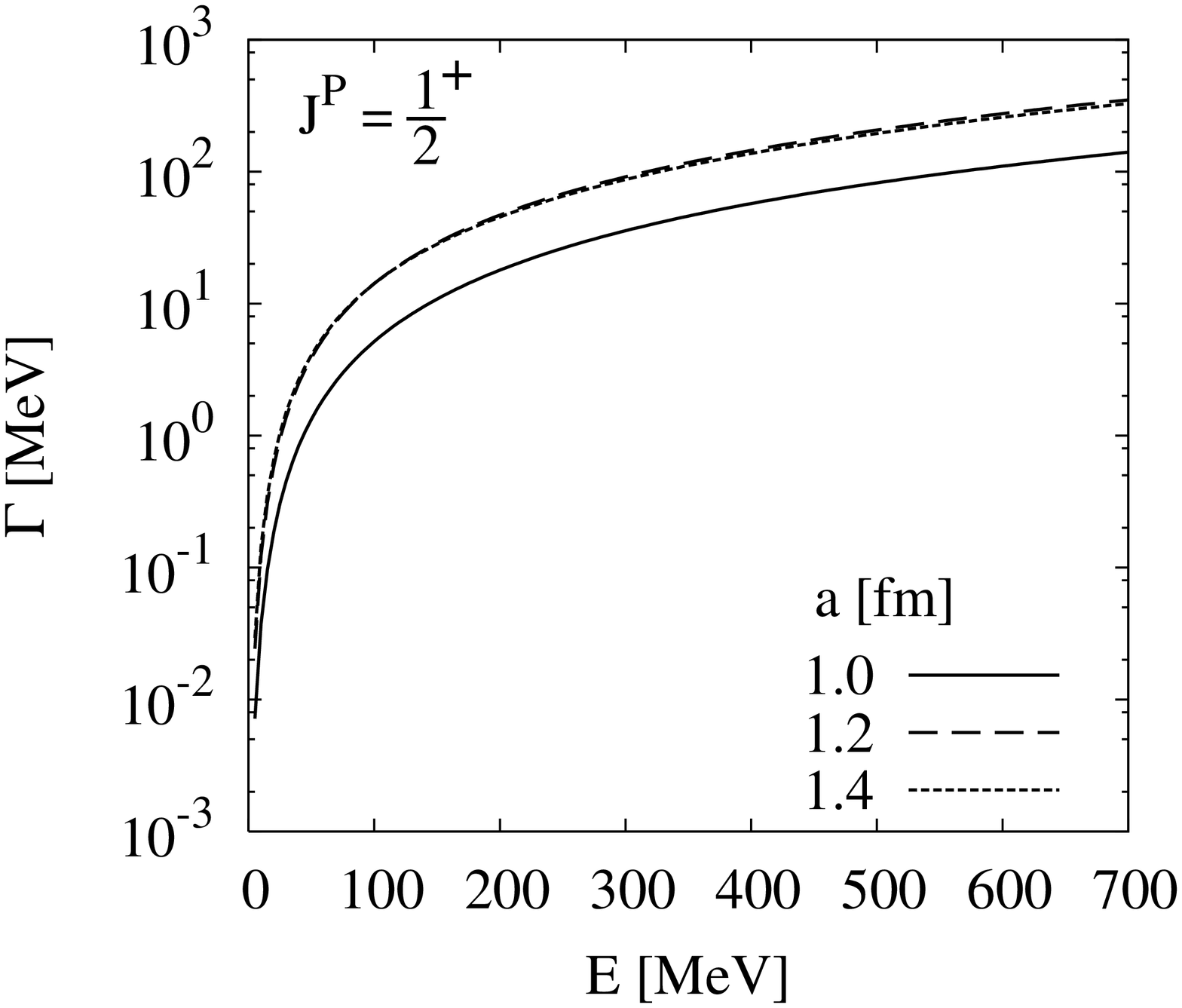}}}
\\
\hspace*{-6.8cm}
\rotatebox{270}{\resizebox{5.5cm}{!}{\includegraphics{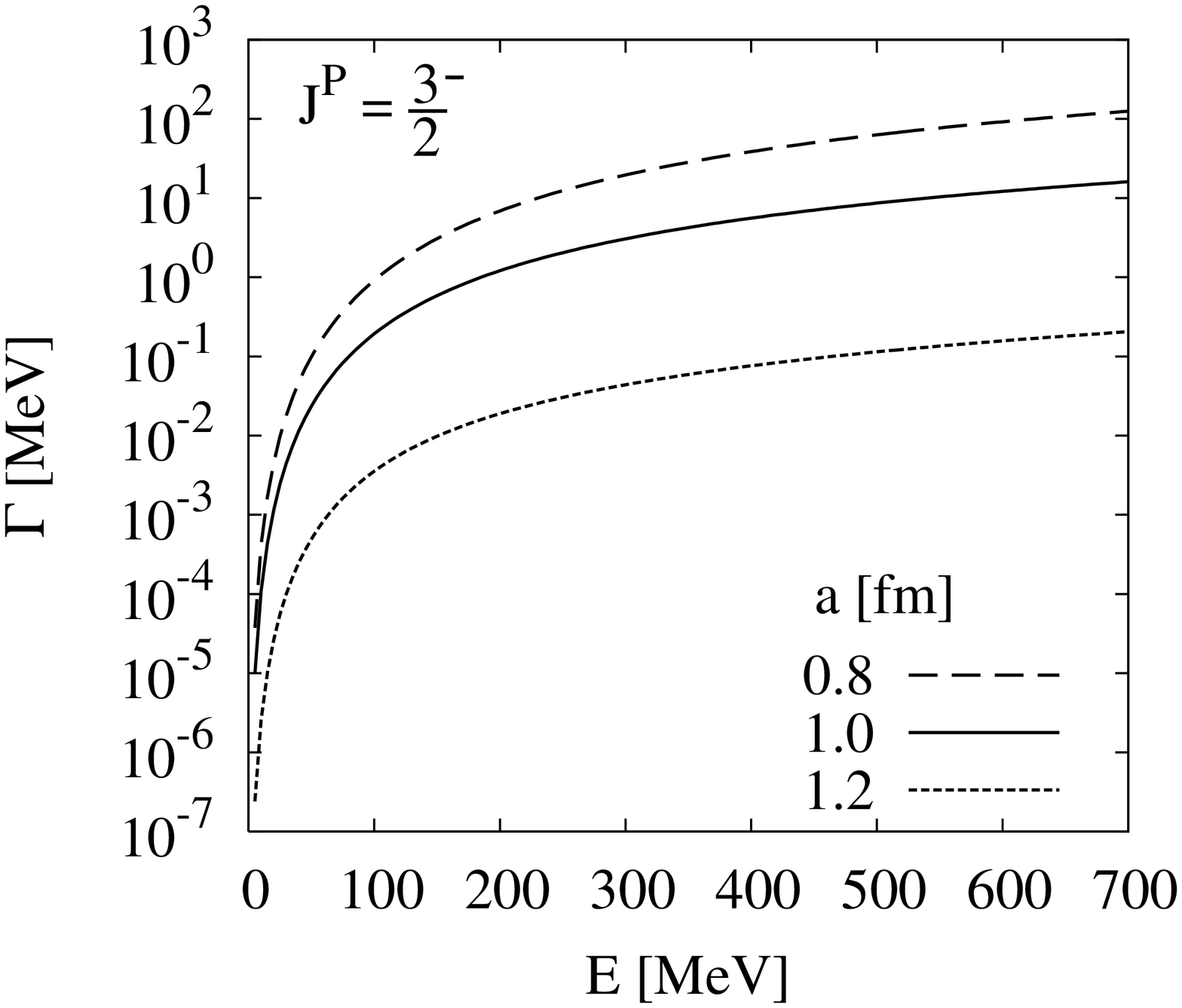}}}
\end{tabular}
\end{center}
\caption{The decay width $\Gamma$ of the $\Theta^+$ to the 
$nK^+$ channel for $J^P=\frac{1}{2}^{\pm}$ and to the 
$nK^{*+}$ channel for $J^P=\frac{3}{2}^{-}$. $E$ is the 
decay energy from the respective threshold 
and $a$ is the channel radius. The AL1 potential is used.}
\label{width.vs.E}
\end{figure}

We estimate the following decay width according to 
Eq.~(\ref{decay.Rmatrix}): the $nK^+$ decay of the $\Theta^+$ with 
$\frac{1}{2}^{\pm}$ and the $nK^{*+}$ decay of the 
$\Theta^+$ with $\frac{3}{2}^{-}$. Because the calculated mass is 
subject to change due to the uncertainty of precise knowledge on the 
interaction between the quarks, we display in 
Fig.~\ref{width.vs.E} the width as a 
function of the decay energy $E$. The channel 
radius $a$ relevant to the decay is estimated as a sum 
of their radii, and it is about 1.1 fm using the values of 
Table~\ref{tab.baryon-meson}. 
The channel radius dependence is also presented around $a\!=\!1$ fm.

First let us accept the calculated mass as it is. Then we 
conclude that no pentaquark states $\Theta^+$ appear around 
1540 MeV, but their masses are expected to be larger than 2 GeV. 
The decay width is then very large, which is different from the 
conclusion of~\cite{hiyama}. Even for the $\frac{1}{2}^+$
state, the width would be in the order of 100 MeV. 
In the case of $J^P\!=\!\frac{3}{2}^-$, 
the mass from the $n\!+\!K^{*+}$ threshold is 270 MeV 
for the AL1 potential 
and 320 MeV for the TS potential, respectively, so that the decay 
width to this channel is a few MeV. The magnitude of 
this width is probably small enough to be observed as a 
sharp resonance.  

Next let us assume that, by shifting the calculated mass spectrum,  
one of the calculated $\Theta^+$ resonances 
corresponds to the observed mass, 1540 MeV, namely 100 MeV 
above the $n\!+\!K^+$ threshold. There are following three 
possibilities, and we estimate the decay width from 
Fig.~\ref{width.vs.E} for each case:
\begin{description}
\item {(1)} {\it If the spin-parity of the $\Theta^+$ is $\frac{1}{2}^-$}, its 
width is several tens to 100 MeV, which is too large to be 
compared to the observation. 
\item {(2)} {\it If the spin-parity of the $\Theta^+$ is $\frac{3}{2}^-$}, 
its $nK^+$ decay width is practically zero. Because of the mass 
spectrum predicted by the present model (see Fig~\ref{mass.spectra}), 
another $\Theta^+$ with $\frac{1}{2}^-$ is expected to appear 
below the $\Theta^+(\frac{3}{2}^-)$ as well and its width is in the order 
of a few MeV. 
\item {(3)} {\it If the spin-parity of the $\Theta^+$ is 
$\frac{1}{2}^+(\frac{3}{2}^+)$}, its $nK^+$ decay 
width is about 10 MeV. Below this resonance other 
$\Theta^+$ states with $\frac{1}{2}^-$ and 
$\frac{3}{2}^-$ appear as quasi-bound states. 
\end{description}
The case (1) is apparently in contradiction to experiment. 
The case (2) may remain as an actual possibility 
because the decay width is practically zero. However, in 
this case another pentaquark state with $J^P=\frac{1}{2}^-$ 
with a small width should also be 
observed near or slightly above the $n\!+\!K^+$ threshold. 
No such experimental information is available at present. 
The case (3) can also be an actual possibility and the decay 
width is in the order of 10 MeV. Though the existence of 
the $\Theta^+$ states with $\frac{1}{2}^-$ and $\frac{3}{2}^-$ are also 
predicted in this case, their energies are below 
the $n\!+\!K^+$ threshold. Thus we expect 
no signal for the existence of such states from the 
type of experiment which uses the $K^+n$ invariant mass 
spectrum for identifying the pentaquark. Therefore, 
the case (3) that the spin-parity of the $\Theta^+(1540)$ 
is $\frac{1}{2}^+$ ($\frac{3}{2}^+$) 
seems not to be in contradiction with the available 
experimental information.

\subsection{Structure}
\label{structure}

To discuss the structure of the $\Theta^+$, it is useful 
to calculate a single-particle density and a 
correlation function between 
the particles. These functions $F$ are all defined as 
\begin{equation}
F({\bfi r}) = \langle\Psi_{JM}^P|\ \delta(\widetilde{\omega}{\bfi x}
-{\bfi r})\ |\Psi_{JM}^P\rangle,
\end{equation}
where $\omega\!=\!(\omega_1,\, \omega_2,\, \omega_3,\, \omega_4)$ 
is a vector 
which is chosen appropriately depending on the quantity of interest.  
For example, the density distribution of the $ud$ quarks from 
the center of mass ${\bfi R}_{\rm c.m.}$ of the system, the quark-quark 
correlation function and the quark-antiquark (${\bar s}$) 
correlation function can be calculated by choosing 
$\omega\!=\!(0, \, 0,\, -\frac{3}{4},\,  
\frac{m_{s}}{4m_{ud}+m_s})$ ($\widetilde{\omega}{\bfi x}\!=\!
{\bfi r}_4\!-\!{\bfi R}_{\rm c.m.}$), $\omega\!=\!(1,\,  0, \, 
0, \, 0)$ ($\widetilde{\omega}{\bfi x}\!=\!{\bfi r}_1\!-\!
{\bfi r}_2$), and $\omega\!=\!(0, \, 0,\, 
-\frac{3}{4}, \, 1)$ ($\widetilde{\omega}{\bfi x}\!
=\!{\bfi r}_4\!-\!{\bfi r}_5$), respectively, because the wave 
function is properly antisymmetrized. The choice of $\omega\!
=\!(0, \, 0, \, 0, \, 1)$ gives the 
distribution ($q^4$-${\bar s}$) of the $\bar s$ from the center 
of mass of the four quarks. When the delta function, 
$\delta(\widetilde{\omega}{\bfi x}-{\bfi r})$, is expanded in multipoles,
only the monopole term contributes to the function $F({\bf r})$ for 
both cases of $J^P\!=\!\frac{1}{2}^{\pm}$, so $F({\bf r})$ becomes spherically
symmetric.

\begin{figure}[b]
\begin{center}
\begin{tabular}{cc}
\rotatebox{270}{\resizebox{5.5cm}{!}{\includegraphics{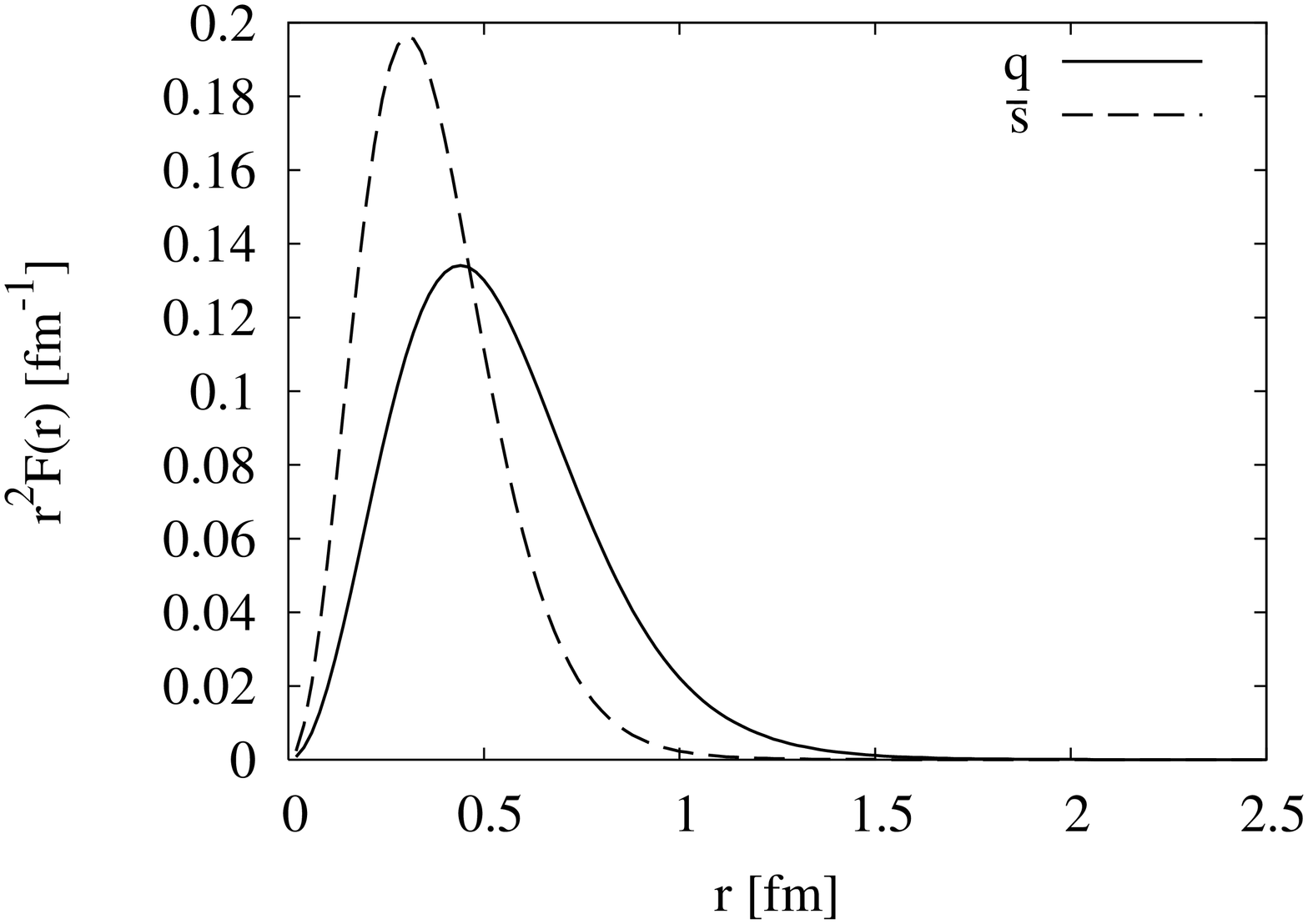}}}
\rotatebox{270}{\resizebox{5.5cm}{!}{\includegraphics{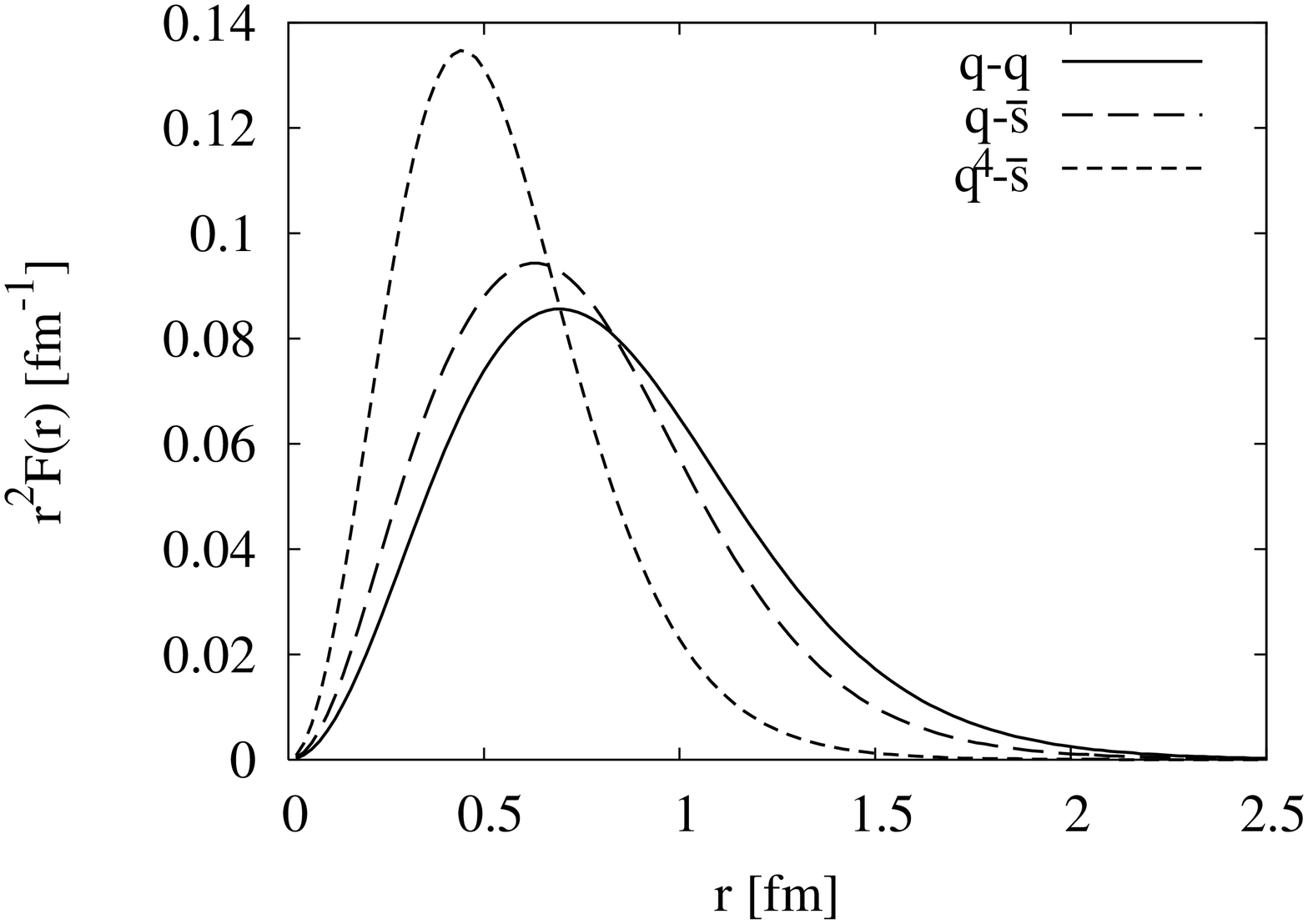}}}
\end{tabular}
\end{center}
\caption{The density distributions of the quarks (left) and the 
correlation functions between the quarks (right) for the $\Theta^+$ 
with $J^P=\frac{1}{2}^-$. The AL1 potential is used.}
\label{density.-}
\end{figure}

\begin{figure}[t]
\begin{center}
\begin{tabular}{cc}
\rotatebox{270}{\resizebox{5.5cm}{!}{\includegraphics{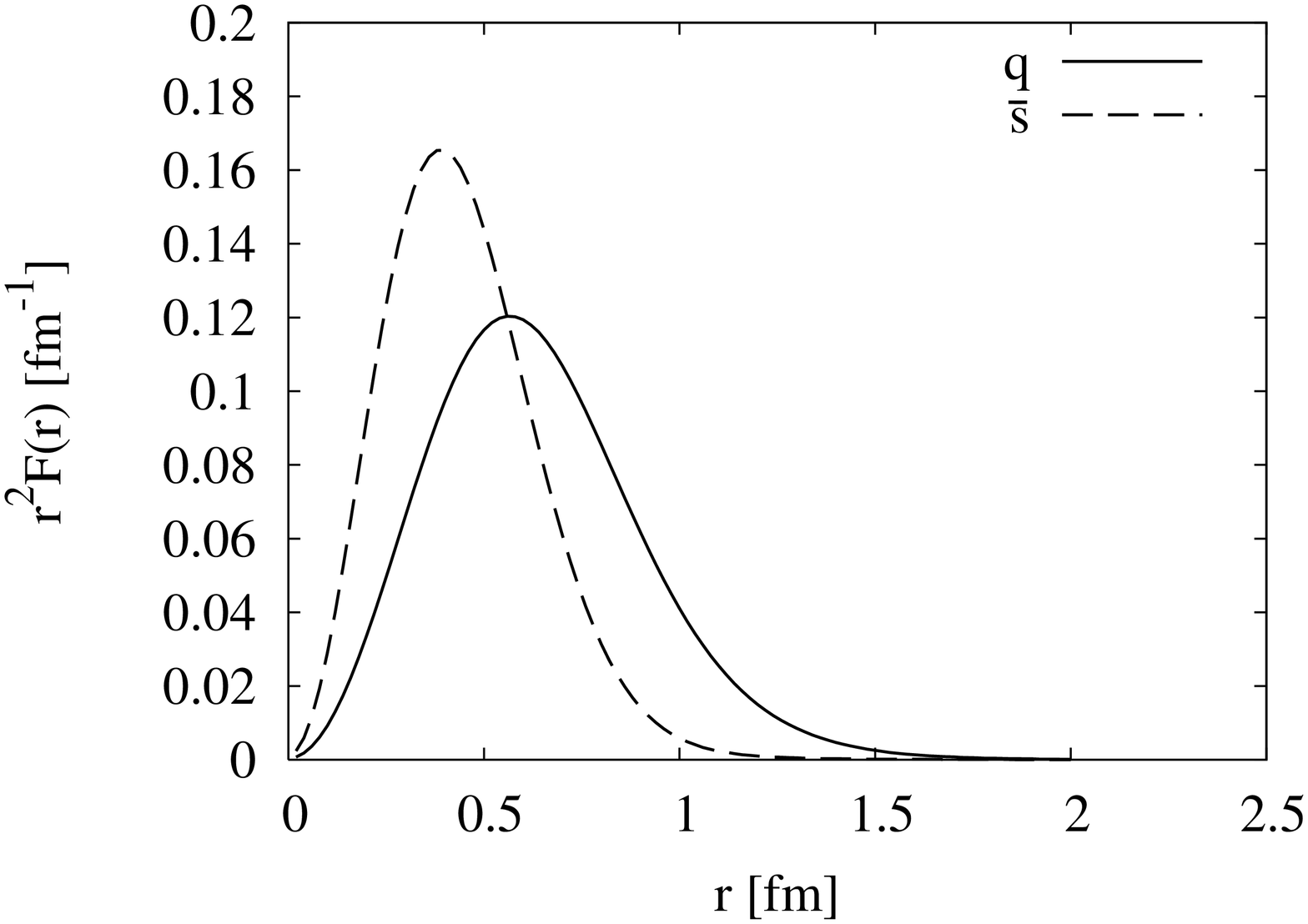}}}
\rotatebox{270}{\resizebox{5.5cm}{!}{\includegraphics{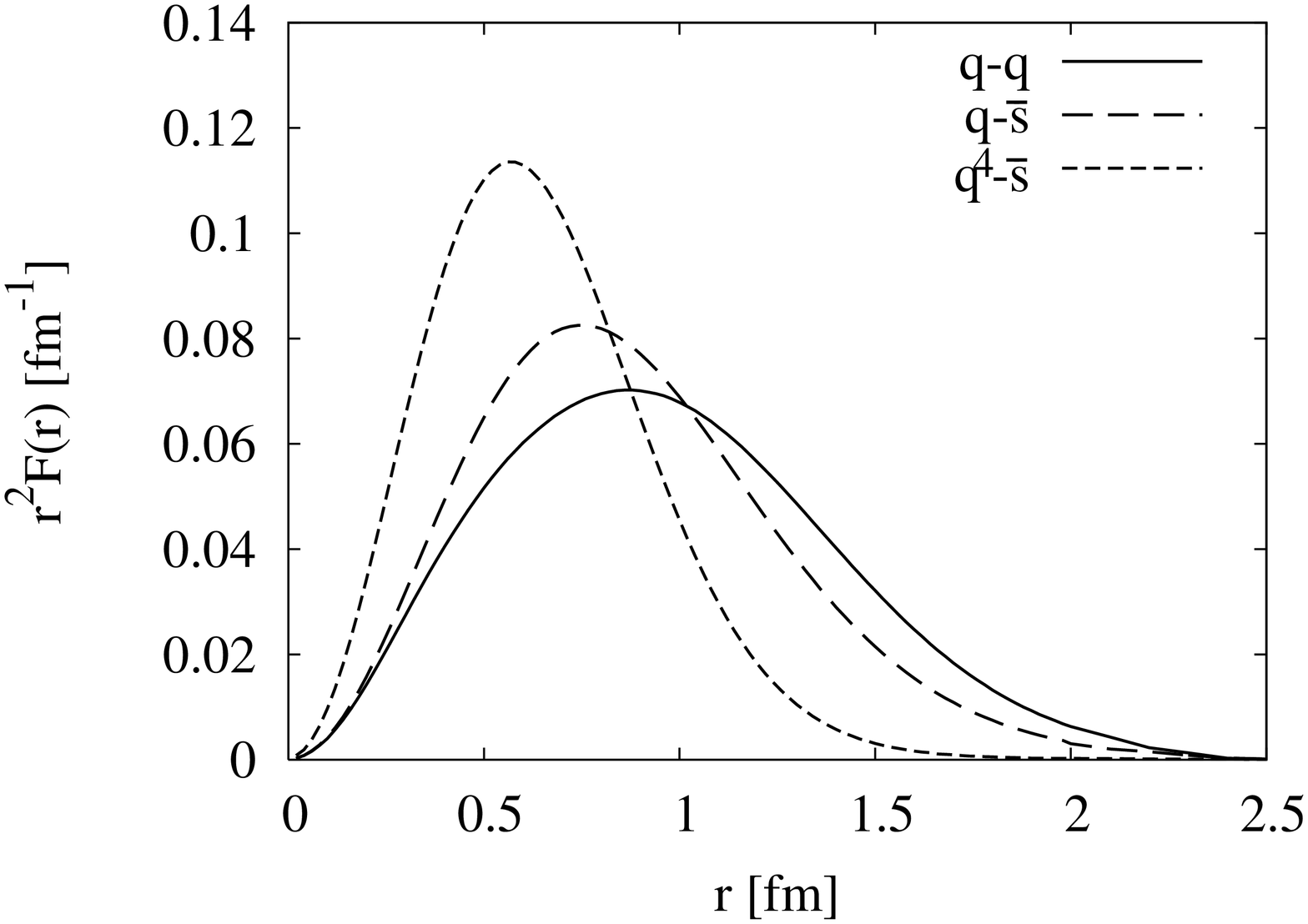}}}
\end{tabular}
\end{center}
\caption{The same as Fig.~\ref{density.-} but for $J^P=\frac{1}{2}^+$. }
\label{density.+}
\end{figure}

Figure~\ref{density.-} displays the density distributions as well 
as the correlation functions for the $\Theta^+$ with 
$J^P\!=\!\frac{1}{2}^-$. It is seen that the distribution 
of the $\bar s$ is 
confined in the smaller region around the center of mass of the 
system than the $ud$-quark distribution. The 
probability density reaches a maximum at about 0.30 fm for $\bar s$ and 
about 0.44 fm for $ud$, respectively. This feature seems to arise 
from the fact that the quark-antiquark interaction is stronger 
than the quark-quark interaction. Corresponding to this feature, 
we see from the correlation function that the average distance 
between the $ud$ and $\bar s$ quarks is shorter than that between 
the $ud$ quarks. Figure~\ref{density.+} displays the similar distributions 
for the $\Theta^+$ with $\frac{1}{2}^+$. The distribution 
extends to larger distances than that of $J^P\!=\!\frac{1}{2}^-$. 
Again, the $\bar s$ distribution has a peak at smaller 
distances from the center of mass than the $ud$ distribution.

Another structure information is obtained by
decomposing the wave function into various channels of the spin, 
isospin and color spaces. To this end, let us define the channel 
wave function $\Phi_c$. See sects.~\ref{wavefunction} and 
\ref{trans.basis}. In the diquark-diquark decomposition $\Phi_c$ is 
defined as 
\begin{eqnarray}
\Phi_{cM_S}&=&\frac{1}{\sqrt{2(1+\delta_{12,34})}}\Bigg\{  
\chi^{\rm DD}_{(S_{12}S_{34}S_{1234})SM_S}\xi^{\rm DD}_{(T_{12}T_{34}0)00}
C^{\rm DD}_{(\Gamma_{12}\Gamma_{34}(10))(00)000}
\nonumber \\
& & \qquad \quad \quad \ + \,\pi \, \sigma \, 
\chi^{\rm DD}_{(S_{34}S_{12}S_{1234})SM_S}\xi^{\rm DD}_{(T_{34}T_{12}0)00}
C^{\rm DD}_{(\Gamma_{34}\Gamma_{12}(10))(00)000}\Bigg\},
\end{eqnarray}
where $\delta_{12,34}\!=\!\delta_{S_{12}S_{34}}\delta_{T_{12}T_{34}}
\delta_{\Gamma_{12}\Gamma_{34}}$, and $\sigma$ is the phase defined by 
\begin{equation}
\sigma=(-1)^{S_{12}+S_{34}-S_{1234}+T_{12}+T_{34}+\lambda_{12}
+\lambda_{34}+\mu_{12}+\mu_{34}-1}.
\end{equation}
The channel index $c$ is characterized by a set of values 
$(S_{12}, T_{12}, \Gamma_{12}, 
S_{34}, T_{34}, \Gamma_{34}, S_{1234}, \pi)$, where $\pi\!=\!\pm 1$ 
determines the parity of the $\Phi_c$. 
The $\Phi_c$ with $\pi\!=\!+1$ is symmetric with respect to 
the simultaneous interchange of the spin, isospin and color 
parts of the two diquarks, and the $\Phi_c$ with 
$\pi\!=\!-1$ is antisymmetric. In the case of 
$\delta_{12,34}\!=\!1$, 
a combination of $\pi \sigma\!=\!1$ only is possible, that is, 
either 
$\pi\!=\!+1$ or $\pi\!=\!-1$ is allowed depending on whether 
$S_{1234}\!=\!1$ or $S_{1234}\!=\!0$.  
In the baryon-meson decomposition $\Phi_c$ is defined as
\begin{equation}
\Phi_{cM_S}=\chi^{\rm BM}_{(S_{12}S_{123}S_{45})SM_S}
\xi^{\rm BM}_{(T_{12}T_{123}\frac{1}{2})00}
C^{\rm BM}_{(\Gamma_{12}\Gamma_{123}\Gamma_{45})(00)000}.
\end{equation}
Here the channel index $c$ stands for $(S_{12}, T_{12}, 
\Gamma_{12}, S_{123}, T_{123}, \Gamma_{123}, S_{45}, \Gamma_{45})$. 
In particular, the $NK$ channel wave function is defined as 
\begin{equation}
\Phi_{NKM_S}=\frac{1}{\sqrt{2}}\Bigg\{
\chi^{\rm BM}_{(0\frac{1}{2}0)\frac{1}{2}M_S}
\xi^{\rm BM}_{(0\frac{1}{2}\frac{1}{2})00}
+\chi^{\rm BM}_{(1\frac{1}{2}0)\frac{1}{2}M_S}
\xi^{\rm BM}_{(1\frac{1}{2}\frac{1}{2})00}\Bigg\}
C^{\rm BM}_{((01)(00)(00))(00)000}.
\end{equation}
The probability $P_c(J^P)$ of finding the channel $c$
in the $\Theta^+$ is calculated as the expectation value 
of the projector $\sum_{M_S}\mid \Phi_{cM_S}\rangle \langle \Phi_{cM_S}\mid$:
\begin{equation}
P_c(J^P)=\sum_{M_S} \langle \Psi_{JM}^P \mid \Phi_{cM_S}\rangle 
\langle \Phi_{cM_S}\mid \Psi_{JM}^P \rangle.
\end{equation}

Values of $P_c(\frac{1}{2}^{\pm})$ are listed in 
Table~\ref{comp.diquark} for the diquark model and in 
Table~\ref{comp.baryon-meson} for the baryon-meson model. 
The channel of the first row in Table~\ref{comp.diquark} represents 
the diquark model of Jaffe-Wilczek~\cite{jaffe}. It is seen 
that this channel occupies a relatively large component in the 
$\frac{1}{2}^+$ state as expected but its magnitude 
is not overwhelmingly 
large. Following this channel, the diquark pair with
$S_{12}T_{12}\Gamma_{12}\!=\!11(01)$ and
$S_{34}T_{34}\Gamma_{34}\!=\!11(20)$ 
occupies a significant weight for $J^P=\frac{1}{2}^+$. 
The channel of the third row corresponds to that of the diquark-triquark 
model~\cite{karliner} as shown in Eq.~(\ref{klmodel}). 
We see that the configuration of the Karliner-Lipkin model 
is dominating in the $\frac{1}{2}^-$ 
state but it is not a main configuration in the $\frac{1}{2}^+$ state. 
With the probability of more than 80 \%, the two diquarks in the 
$\frac{1}{2}^-$ state are found in either the Karliner-Lipkin 
channel or the channel with $S_{12}T_{12}\Gamma_{12}\!=\!11(01),
S_{34}T_{34}\Gamma_{34}\!=\!11(01), S_{1234}\!=\!1$. 
It is seen from Table~\ref{comp.baryon-meson} that the components 
in the baryon-meson decomposition are spread over many channels. 
Relatively large components do not 
necessarily appear in colorless baryons. The probability summed over 
the colored baryon-meson channels is 67 \%, which signals the 
importance of the hidden color components. 
The $NK$ component, $P_{NK}(J^{P})$, 
is calculated to be 0.14 for the $\frac{1}{2}^-$ state and  
0.073 for the $\frac{1}{2}^+$ state, respectively. 
The magnitude of this channel is related to the reduced width 
to the $nK^+$ channel discussed 
in sect.~\ref{decay}, and it is certainly small, especially for 
the $\frac{1}{2}^+$ state.

\begin{table}[t]
\caption{Decompositions of the $\Theta^+$ resonances 
into the spin, isospin and color channels of the diquark model. 
See Tables~\ref{tab.spin} and \ref{tab.color} for the channel labels. 
$T_{1234}\!=\!0$ and $\Gamma_{1234}\!=\!(10)$ are abbreviated. 
$\pi$ is the parity of the channel wave function with respect to the 
diquark-diquark exchange. The AL1 potential is used. }
\begin{center}
\begin{tabular}{cccccc}
\hline\hline
$S_{12} T_{12} \Gamma_{12}$ & $S_{34} T_{34} \Gamma_{34}$ &
\ $S_{1234}$ & $\pi$ & \ $P_c(\frac{1}{2}^-)$ 
& \ $P_c(\frac{1}{2}^+)$ \\
\hline
\ 0\ \  0\ \  (01) & \ 0\ \  0\ \  (01) & 0 & $-1$ & 0.0005 & 0.2067\\
\ 0\ \  0\ \  (01) & \ 1\ \  0\ \  (20) & 1 & $-1$ & 0.0017 & 0.0457\\
\ 0\ \  0\ \  (01) & \ 1\ \  0\ \  (20) & 1 & $+1$ & 0.4933 & 0.0272\\
\ 0\ \  1\ \  (01) & \ 1\ \  1\ \  (01) & 1 & $-1$ & 0.0008 & 0.0316\\
\ 0\ \  1\ \  (01) & \ 1\ \  1\ \  (01) & 1 & $+1$ & 0.0008 & 0.0316\\
\ 0\ \  1\ \  (20) & \ 1\ \  1\ \  (01) & 1 & $-1$ & 0.0012 & 0.0310\\
\ 0\ \  1\ \  (20) & \ 1\ \  1\ \  (01) & 1 & $+1$ & 0.1649 & 0.0126\\
\ 1\ \  1\ \  (01) & \ 1\ \  1\ \  (01) & 0 & $-1$ & 0.0003 & 0.0062\\
\ 1\ \  1\ \  (01) & \ 1\ \  1\ \  (01) & 1 & $+1$ & 0.3290 & 0.0215\\
\ 0\ \  0\ \  (01) & \ 0\ \  0\ \  (20) & 0 & $-1$ & 0.0002 & 0.0149\\
\ 0\ \  0\ \  (01) & \ 0\ \  0\ \  (20) & 0 & $+1$ & 0.0002 & 0.0149\\
\ 0\ \  0\ \  (01) & \ 1\ \  0\ \  (01) & 1 & $-1$ & 0.0007 & 0.0150\\
\ 0\ \  0\ \  (01) & \ 1\ \  0\ \  (01) & 1 & $+1$ & 0.0007 & 0.0150\\
\ 0\ \  1\ \  (01) & \ 1\ \  1\ \  (20) & 1 & $-1$ & 0.0000 & 0.0123\\
\ 0\ \  1\ \  (01) & \ 1\ \  1\ \  (20) & 1 & $+1$ & 0.0006 & 0.0068\\
\ 0\ \  1\ \  (01) & \ 0\ \  1\ \  (20) & 0 & $-1$ & 0.0001 & 0.0415\\
\ 0\ \  1\ \  (01) & \ 0\ \  1\ \  (20) & 0 & $+1$ & 0.0001 & 0.0415\\
\ 1\ \  1\ \  (01) & \ 1\ \  1\ \  (20) & 0 & $-1$ & 0.0003 & 0.1151\\
\ 1\ \  1\ \  (01) & \ 1\ \  1\ \  (20) & 0 & $+1$ & 0.0003 & 0.1151\\
\ 1\ \  0\ \  (01) & \ 1\ \  0\ \  (20) & 0 & $-1$ & 0.0001 & 0.0415\\
\ 1\ \  0\ \  (01) & \ 1\ \  0\ \  (20) & 0 & $+1$ & 0.0001 & 0.0415\\
\ 1\ \  1\ \  (01) & \ 1\ \  1\ \  (20) & 1 & $-1$ & 0.0008 & 0.0166\\
\ 1\ \  1\ \  (01) & \ 1\ \  1\ \  (20) & 1 & $+1$ & 0.0008 & 0.0166\\
\ 1\ \  0\ \  (01) & \ 1\ \  0\ \  (20) & 1 & $-1$ & 0.0006 & 0.0299\\
\ 1\ \  0\ \  (01) & \ 1\ \  0\ \  (20) & 1 & $+1$ & 0.0006 & 0.0299\\
\ 0\ \  1\ \  (01) & \ 0\ \  1\ \  (01) & 0 & $-1$ & 0.0000 & 0.0000\\
\ 0\ \  0\ \  (20) & \ 1\ \  0\ \  (01) & 1 & $-1$ & 0.0000 & 0.0041\\
\ 0\ \  0\ \  (20) & \ 1\ \  0\ \  (01) & 1 & $+1$ & 0.0005 & 0.0080\\
\ 1\ \  0\ \  (01) & \ 1\ \  0\ \  (01) & 0 & $-1$ & 0.0000 & 0.0000\\
\ 1\ \  0\ \  (01) & \ 1\ \  0\ \  (01) & 1 & $+1$ & 0.0006 & 0.0057\\
\hline
           &            &   &   &  1.0000 & 1.0000\\
\hline\hline
\end{tabular}
\end{center}
\label{comp.diquark}
\end{table}

\begin{table}[b]
\caption{Decompositions of the $\Theta^+$ resonances 
into the spin, isospin and color channels of the baryon-meson 
model. See Tables~\ref{tab.spin} and \ref{tab.color} 
for the channel labels. The AL1 potential is used. }
\begin{center}
{\small 
\begin{tabular}{ccccc}
\hline\hline
$S_{12}T_{12}\Gamma_{12}$ & \ $S_{123}T_{123}\Gamma_{123}$ & 
\ $S_{45}T_{45}\Gamma_{45}$ 
& \ $P_c(\frac{1}{2}^-)$ 
& \ $P_c(\frac{1}{2}^+)$ \\
\hline
\ 0\ \ 0\ \ (01) & \ \ $\frac{1}{2}\ \ \frac{1}{2}\ \ (00)$ 
& \ \ 0\ \ $\frac{1}{2}$\ \  (00) & 0.0022 &  0.0588 \\
\ 0\ \ 0\ \ (01) & \ \ $\frac{1}{2}\ \ \frac{1}{2}\ \ (00)$ 
& \ \ 1\ \ $\frac{1}{2}$\ \ (00) & 0.0011 &  0.1615 \\
\ 1\ \ 1\ \ (01) & \ \ $\frac{1}{2}\ \ \frac{1}{2}\ \ (00)$ 
& \ \ 0\ \ $\frac{1}{2}$\ \ (00) & 0.1529 &  0.0179 \\
\ 1\ \ 1\ \ (01) & \ \ $\frac{1}{2}\ \ \frac{1}{2}\ \ (00)$ 
& \ \ 1\ \ $\frac{1}{2}$\ \ (00) & 0.0512 &  0.0124 \\
\ 0\ \ 1\ \ (01) & \ \ $\frac{1}{2}\ \ \frac{1}{2}\ \ (00)$ 
& \ \ 0\ \ $\frac{1}{2}$\ \ (00) & 0.0006 &  0.0222 \\
\ 0\ \ 1\ \ (01) & \ \ $\frac{1}{2}\ \ \frac{1}{2}\ \ (00)$ 
& \ \ 1\ \ $\frac{1}{2}$\ \ (00) & 0.0002 &  0.0095 \\
\ 1\ \ 0\ \ (01) & \ \ $\frac{1}{2}\ \ \frac{1}{2}\ \ (00)$ 
& \ \ 0\ \ $\frac{1}{2}$\ \ (00) & 0.0005 &  0.0066 \\
\ 1\ \ 0\ \ (01) & \ \ $\frac{1}{2}\ \ \frac{1}{2}\ \ (00)$ 
& \ \ 1\ \ $\frac{1}{2}$\ \ (00) & 0.0002 &  0.0028 \\
\ 0\ \ 0\ \ (01) & \ \ $\frac{1}{2}\ \ \frac{1}{2}\ \ (11)$ 
& \ \ 0\ \ $\frac{1}{2}$\ \ (11) & 0.1841 &  0.0352 \\
\ 0\ \ 1\ \ (20) & \ \ $\frac{1}{2}\ \ \frac{1}{2}\ \ (11)$ 
& \ \ 0\ \ $\frac{1}{2}$\ \ (11) & 0.0623 &  0.0267 \\
\ 1\ \ 0\ \ (20) & \ \ $\frac{1}{2}\ \ \frac{1}{2}\ \ (11)$ 
& \ \ 0\ \ $\frac{1}{2}$\ \ (11) & 0.0622 &  0.0266 \\
\ 1\ \ 1\ \ (01) & \ \ $\frac{1}{2}\ \ \frac{1}{2}\ \ (11)$ 
& \ \ 0\ \ $\frac{1}{2}$\ \ (11) & 0.0330 &  0.0642 \\
\ 0\ \ 0\ \ (01) & \ \ $\frac{1}{2}\ \ \frac{1}{2}\ \ (11)$ 
& \ \ 1\ \ $\frac{1}{2}$\ \ (11) & 0.0615 &  0.0175 \\
\ 0\ \ 1\ \ (20) & \ \ $\frac{1}{2}\ \ \frac{1}{2}\ \ (11)$ 
& \ \ 1\ \ $\frac{1}{2}$\ \ (11) & 0.0209 &  0.0365 \\
\ 1\ \ 0\ \ (20) & \ \ $\frac{1}{2}\ \ \frac{1}{2}\ \ (11)$ 
& \ \ 1\ \ $\frac{1}{2}$\ \ (11) & 0.0209 &  0.0470 \\
\ 1\ \ 1\ \ (01) & \ \ $\frac{1}{2}\ \ \frac{1}{2}\ \ (11)$ 
& \ \ 1\ \ $\frac{1}{2}$\ \ (11) & 0.0113 &  0.0701 \\
\ 0\ \ 0\ \ (20) & \ \ $\frac{1}{2}\ \ \frac{1}{2}\ \ (11)$ 
& \ \ 0\ \ $\frac{1}{2}$\ \ (11) & 0.0002 &  0.0082 \\
\ 0\ \ 1\ \ (01) & \ \ $\frac{1}{2}\ \ \frac{1}{2}\ \ (11)$ 
& \ \ 0\ \ $\frac{1}{2}$\ \ (11) & 0.0002 &  0.0190 \\
\ 1\ \ 0\ \ (01) & \ \ $\frac{1}{2}\ \ \frac{1}{2}\ \ (11)$ 
& \ \ 0\ \ $\frac{1}{2}$\ \ (11) & 0.0004 &  0.0347 \\
\ 1\ \ 1\ \ (20) & \ \ $\frac{1}{2}\ \ \frac{1}{2}\ \ (11)$ 
& \ \ 0\ \ $\frac{1}{2}$\ \ (11) & 0.0006 &  0.0201 \\
\ 0\ \ 0\ \ (20) & \ \ $\frac{1}{2}\ \ \frac{1}{2}\ \ (11)$ 
& \ \ 1\ \ $\frac{1}{2}$\ \ (11) & 0.0002 &  0.0127 \\
\ 0\ \ 1\ \ (01) & \ \ $\frac{1}{2}\ \ \frac{1}{2}\ \ (11)$ 
& \ \ 1\ \ $\frac{1}{2}$\ \ (11) & 0.0002 &  0.0319 \\
\ 1\ \ 0\ \ (01) & \ \ $\frac{1}{2}\ \ \frac{1}{2}\ \ (11)$ 
& \ \ 1\ \ $\frac{1}{2}$\ \ (11) & 0.0002 &  0.0281 \\
\ 1\ \ 1\ \ (20) & \ \ $\frac{1}{2}\ \ \frac{1}{2}\ \ (11)$ 
& \ \ 1\ \ $\frac{1}{2}$\ \ (11) & 0.0004 &  0.1092 \\
\ 1\ \ 1\ \ (01) & \ \ $\frac{3}{2}\ \ \frac{1}{2}\ \ (00)$ 
& \ \ 1\ \ $\frac{1}{2}$\ \ (00) & 0.1239 &  0.0299 \\
\ 1\ \ 0\ \ (20) & \ \ $\frac{3}{2}\ \ \frac{1}{2}\ \ (11)$ 
& \ \ 1\ \ $\frac{1}{2}$\ \ (11) & 0.1652 &  0.0343 \\
\ 1\ \ 1\ \ (01) & \ \ $\frac{3}{2}\ \ \frac{1}{2}\ \ (11)$ 
& \ \ 1\ \ $\frac{1}{2}$\ \ (11) & 0.0419 &  0.0185 \\
\ 1\ \ 0\ \ (01) & \ \ $\frac{3}{2}\ \ \frac{1}{2}\ \ (00)$ 
& \ \ 1\ \ $\frac{1}{2}$\ \ (00) & 0.0006 &  0.0118 \\
\ 1\ \ 0\ \ (01) & \ \ $\frac{3}{2}\ \ \frac{1}{2}\ \ (11)$ 
& \ \ 1\ \ $\frac{1}{2}$\ \ (11) & 0.0004 &  0.0141 \\
\ 1\ \ 1\ \ (20) & \ \ $\frac{3}{2}\ \ \frac{1}{2}\ \ (11)$ 
& \ \ 1\ \ $\frac{1}{2}$\ \ (11) & 0.0004 &  0.0119 \\
\hline
& & &  1.0000 & 1.0000  \\
\hline\hline
\end{tabular}
}
\end{center}
\label{comp.baryon-meson}
\end{table}

\section{Summary}
\label{summary}

We have studied the mass and the decay width of the $\Theta^+$ 
consisting of $uudd{\bar s}$ quarks in the constituent quark model. 
The quarks are assumed to interact via a variant of the 
one-gluon exchange potential plus a phenomenological confinement 
potential which reproduce the masses of e.g., $N$, 
$K$ and $K^*$ reasonably well. 
The five-quark states are described using the correlated Gaussian 
basis together with the general spin-isospin-color wave functions. 
The basis functions are set up with the stochastic variational 
method. One of the advantages of the present approach is that  
the importance of all the possible color configurations can be 
tested, so that 
the hidden color components are naturally taken into account 
in the calculation. We have performed coupled-channels calculations 
to predict the $\Theta^+$ masses 
for $J^P\!=\!\frac{1}{2}^-\, (L\!=\!0,S\!=\!\frac{1}{2},T\!=\!0)$, 
$\frac{1}{2}^+(\frac{3}{2}^+)\, (L\!=\!1,S\!=\!\frac{1}{2},T\!=\!0)$ 
and $\frac{3}{2}^-\, (L\!=\!0,S\!=\!\frac{3}{2},T\!=\!0)$. 
The $NK$ and $NK^*$ channels are explicitly included as they are 
expected to be important to obtain the decay width. 
The real stabilization method is successfully used to locate the 
resonance in the continuum states above the $N\!+\!K$ threshold. 
The ground state is found to be $\frac{1}{2}^-$ and its mass is 
about 2 GeV, which is 500-600 MeV higher than 
the observed value. The first and second excited states are 
$\frac{3}{2}^-$ and $\frac{1}{2}^+(\frac{3}{2}^+)$. 
The color magnetic and kinetic energy terms of the Hamiltonian 
are the most important pieces which contribute to this level ordering. 

To estimate the decay width of the $\Theta^+$, 
we have used the $R$-matrix theory in which most crucial is 
the reduced width amplitude for the relevant decay channel. 
The decay channel is specified by the angular momentum $I$, the 
addition of the spins of the baryon and the meson, as well as 
the relative orbital angular momentum $\ell$ between them. The 
calculated decay width is the $nK^+$ decay with $(I,\ell)\! =\! 
(\frac{1}{2},0)$ for $\frac{1}{2}^-$ and  
$(I,\ell)\! = \!(\frac{1}{2}, 1)$ for $\frac{1}{2}^+
(\frac{3}{2}^+)$, respectively. For the $\Theta^+$ with 
$J^P\!=\!\frac{3}{2}^-$, the decay to the $nK^+$ channel is 
forbidden because the corresponding reduced width amplitude vanishes, 
so the decay width to the $nK^{*+}$ channel is calculated. All the 
reduced width amplitudes are small compared to the Wigner limit value. 
As the mass 
of the $\Theta^+$ is very large compared to the threshold, the decay 
width becomes very broad except for the $\frac{3}{2}^-$ state whose 
decay width to the $nK^{*+}$ channel is only a few MeV. 
We have also estimated the decay width by shifting the calculated mass 
to the observed mass, i.e., 100 MeV above the $n\!+\!K$ threshold. This 
shift causes no change in the resonance wave functions. 
The possibility that the observed state is $\frac{1}{2}^-$ is ruled 
out because its width is still too large. The case that it 
is either $\frac{3}{2}^-$ or $\frac{1}{2}^+ (\frac{3}{2}^+)$ appears 
not to be inconsistent with the observation of the small width. 
However, in both cases other pentaquark state with different $J^P$ 
is expected to exist below the observed one at 1540 MeV. Its 
experimental confirmation will give us a support for 
the $\Theta^+$. 

The structure of the $\frac{1}{2}^\pm$ states is discussed through 
the density distributions of $ud$ and ${\bar s}$ quarks as well 
as the two-particle correlation functions of the quarks. It is 
found that the $\bar{s}$ quark has a narrower distribution 
near the center of mass of the $uudd$ quarks than the $ud$ quark. 
The average $q$-$q$ distance is longer 
compared to that of $q$-$\bar{s}$. These characteristics can be 
understood from the fact that the 
$q$-$\bar{q}$ interaction is stronger than the 
$q$-$q$ interaction. We have decomposed the resonance wave function 
into various components using the diquark-diquark model and the 
baryon-meson model.  The baryon-meson configuration 
of the Karliner-Lipkin model is in fact a main component 
in the $\frac{1}{2}^-$ state but not an important component in the 
$\frac{1}{2}^+$ state, differing from its 
expectation. In contrast to this, the diquark-diquark 
configuration of the Jaffe-Wilczek model occupies 
the largest 
component in the $\frac{1}{2}^+$ state as expected, but its 
magnitude is not very large. The components are actually spread over many 
diquark-diquark channels.

\begin{acknowledgments}
This work was in part supported by a Grant-in-Aid for Scientific 
Research (No. 14540249) of the Japan Society for the Promotion of 
Science and a Grant for Promotion of Niigata University Research
Projects (2005-2007).
\end{acknowledgments}

\end{document}